\documentclass[runningheads,natbib]{svjour2}
\bibpunct{[}{]}{;}{n}{}{,} 
\smartqed  
\usepackage{graphicx}
\usepackage{aas_macros}
\usepackage{lscape}
\usepackage{psfig}
\def\pdot {\dot P}

\def\ltsima{$\; \buildrel < \over \sim \;$}
\def\lsim{\lower.5ex\hbox{\ltsima}}
\def\gtsima{$\; \buildrel > \over \sim \;$}
\def\gsim{\lower.5ex\hbox{\gtsima}}
\def\msun{~M_{\odot}}
\def\rsun{~R_{\odot}}

\def\qui {1E 1547-54}
\def\uu {4U~0142$+$61}
\def\oo {1E~1048$-$59}
\def\kes {1E~1841$-$045}
\def\axj {AX~J1845$-$02}
\def\rxs {1RXS~J1708$-$40}
\def\ee {1E~2259$+$586}
\def\xte {XTE~J1810$-$197}
\def\cxo {CXOU~J1647$-$45}
\def\smc {CXOU~J0100-72}
\def\zerosei {SGR~1806$-$20}
\def\zerozero {SGR~1900$+$14}
\def\sedici {SGR~1627$-$41}
\def\lmc {SGR~0526$-$66}
\def\quil {1E 1547.0-5408}
\def\uul {4U~0142$+$61}
\def\ool {1E~1048$-$586}
\def\kesl {1E~1841$-$045}
\def\axjl {AX~J1844.8$-$0256}
\def\rxsl {1RXS~J170849$-$400910}
\def\eel {1E~2259$+$586}
\def\xtel {XTE~J1810$-$197}
\def\cxol {CXOU~J164710.2$-$455216}
\def\smcl {CXOU~J010043.1-721134}

\journalname{Astronomy and Astrophysics Review}
\begin{document}

\title{The strongest cosmic magnets:  Soft Gamma-ray Repeaters and Anomalous X-ray Pulsars }

\titlerunning{Soft Gamma-ray Repeaters and Anomalous X-ray Pulsars}        

\author{Sandro Mereghetti}


\institute{S. Mereghetti \at
              INAF - IASF Milano \\
              via E.Bassini 15, I-20133 Milano, Italy\\
              \email{sandro@iasf-milano.inaf.it}           
}

\date{Received: date}

\maketitle

\begin{abstract}
Two classes of X-ray pulsars, the Anomalous X--ray Pulsars and the
Soft Gamma--ray Repeaters, have been recognized in the last decade
as the most promising candidates for being magnetars: isolated
neutron stars powered by magnetic energy. I review the
observational properties of these objects, focussing on the most
recent results, and their interpretation in the magnetar model.
Alternative explanations, in particular those based on accretion
from residual disks, are also considered. The possible relations
between these sources  and other classes of neutron stars and
astrophysical objects are also discussed.
\end{abstract}

\section{Introduction}
\label{intro}

Magnetars are neutron stars with magnetic fields much larger than
the quantum critical value $B_{QED}=\frac{m^2 c^3}{\hbar
e}$=4.4$\times10^{13}$ G, at which the energy between Landau
levels of electrons equals their rest mass.  Their magnetic fields
are at least 100-1000 times stronger than those of the typical
neutron stars observed as radio pulsars powered by the loss of
rotational energy, or shining in X-rays thanks to the accretion of
matter from binary companion stars. Magnetic field is the ultimate
energy source of all the observed emission from magnetars
\citep{tho95,tho96}.

Magnetars have attracted   increasing attention in the last
decade, being extremely interesting objects, both from the
physical and astronomical point of view. They allow us to observe
and study several phenomena taking place in magnetic field
conditions not available elsewhere (see, e.g., \citep{har06}).
Their astrophysical importance is due to the fact that they
broadened our views of how neutron stars are formed and evolve.
Together with other new classes of neutron stars observed through
the whole electromagnetic spectrum, they indicate that the
classical radio pulsars discovered forty years ago are just one of
the diverse manifestations of neutron stars.

Magnetars are historically divided in two classes of neutron stars
that were independently discovered through  different
manifestations of their high-energy emission: the Soft Gamma-ray
Repeaters (SGRs) and the Anomalous X-ray Pulsars (AXPs). SGRs were
discovered through the detection of short bursts in the hard
X-ray/soft gamma-ray range, and initially considered as a subclass
of gamma-ray bursts \citep{lar86,att87}. AXPs were first detected
in the soft X-ray range ($<$10 keV) and thought to belong to the
population of galactic accreting binaries; only as more X--ray
data accumulated, and deeper optical/IR searches excluded the
presence of bright companion stars, their peculiar properties
started to appear, leading to their classification as a separate
class of pulsars \citep{mer95}. Observations performed over the
last few years led to new discoveries pointing out  many
similarities between these two classes of objects \citep{woo06}.
Thus, the magnetar model initially developed to explain the
extreme properties of the SGRs, difficult to interpret in other
models \citep{dun92}, was  applied also to the AXPs \citep{tho96}.

The main observational properties that led to the recognition of
the AXPs as a homogenous class, different from the more common
accretion-powered pulsars in massive X-ray binaries, were the
following \citep{mer95,mer02}:


\textit{(a)} lack of evidence of binary companions

\textit{(b)} luminosity larger than the spin-down power

\textit{(c)} spin period in the 5-12 s range

\textit{(d)} secular spin-down on timescales of 10$^3$-10$^5$
years

\textit{(e)} no (or very small) long term variability

\textit{(f)} soft X-ray spectrum

\textit{(g)} absence of radio emission

\textit{(h)} (in some cases) association with supernova remnants


\noindent
When the persistent X-ray counterparts of SGRs were
found, it was apparent that they shared many of these properties:
they showed luminosities, periods and period derivatives similar
to the AXPs, but had generally harder spectra. Possible
associations with SNRs were reported for all the four confirmed
SGRs.

After more of ten years of extensive observations in many
wavelengths,  most of the above properties have been consolidated
on the basis of better data, but a few of them (e.g. \textit{(e)}
and \textit{(g)}), sometimes unexpectedly, have not been
confirmed:

\textit{(a)} and \textit{(b)} -- these two properties remain
prerequisite characteristics to exclude more conventional
explanations for newly discovered X-ray pulsars. Much progress has
been done in the search for optical/IR counterparts
(sect.~\ref{optical}) and the resulting faintness of the
candidates has confirmed that standard binary systems powered by
accretion from a companion are excluded.

\textit{(c)} and \textit{(d)} -- the characteristic P and $\pdot$
values of these objects have been confirmed. The reason for the
narrow distribution of period values is not obvious
(sect.~\ref{periods}). The timing signatures have provided a
wealth of important information, through the measurement of noise
and glitches (sect.~\ref{glit}), as well as through the
observation of quasi-periodic oscillations (QPOs) and other
effects during the SGR giant flares (sect.~\ref{gf}).

\textit{(e)} -- one of the most interesting results of the
observations carried out in the last few years is that, at
variance with most manifestations of isolated neutron stars
powered by rotational energy or residual heat, the
magnetically-powered X--ray emission from AXPs and SGRs is
variable on different time scales (sect.~\ref{variab}). Long term
flux variations have now been observed in virtually all objects
for which accurate measurements are available. In addition, there
are a few remarkable cases of transient magnetars, spanning a
range of 2-3 orders of magnitude in luminosity
(sect.~\ref{trans}). On the shortest time scales, the rapid bursts
that were the defining characteristic of SGRs have now been seen
in also in most AXPs, although with smaller peak luminosity and
possibly slightly different properties (sect.~\ref{burst}). The
spectacular flares seen in SGRs (sect.~\ref{gf}) were
traditionally classified in giant and intermediate, but as more
events are found, including those seen in AXPs, it seems that they
rather span a continuum of intensities. A coherent picture
relating all these variability phenomena has not emerged yet. In
several cases there is evidence that the luminosity variations on
medium and long term are associated to sudden events like bursts
or glitches (sect.~\ref{glit}). On the other hand, there are also
long term variations apparently unrelated to such events, although
the sparse coverage of the observations does not allow to draw
firm conclusions.

\textit{(f)} -- the softness of AXP spectra below 10 keV has been
confirmed, but  observations with the INTEGRAL satellite above 20
keV have unexpectedly shown the presence of a significant flux of
hard X-rays in the persistent (i.e. not bursting) emission from
several AXPs and SGRs (sect.~\ref{hardx}). This discovery is
particularly important since it turns out that the bolometric
output from these objects can be dominated by non-thermal
magnetospheric emission.

\textit{(g)} -- another rather unexpected result is the discovery
of pulsed radio emission from two AXPs (sect.~\ref{radio}). This
property seems to be a prerogative of transient magnetars. The
presence of pulsed radio emission, besides its intrinsic interest,
provides a new important diagnostic tool for several other aspects
of the study of magnetars: it allows to derive independent
distance estimates and very precise position determinations,
possibly leading to proper motion measurements. Furthermore, pulse
timing measurements in the radio band can be carried with a higher
precision and on shorter time scales than in X-rays, thus offering
a better tool to study glitches and torque variations.

\textit{(h)} -- the association with SNRs is robust in two or
three objects, but not considered significant in several other
cases that were proposed in the past.

Table~\ref{tab-list} lists all the known magnetars\footnote{I will
use the term magnetar when referring to both AXPs and SGRs.} and
candidate magnetars. Compared to a few years ago, when these
objects were studied mostly in X--rays, it is striking to see the
important role now played by multi-wavelength observations.

Several reviews on magnetars are already available
\citep{mer02b,woo06,kas07}, therefore I will concentrate here
mainly on the more recent developments in this very dynamical
field. In the next three sections I describe the observational
properties of AXPs and SGRs. The main concepts of the magnetar
model are then discussed (sect.~\ref{magnetar}), while alternative
models are presented in section \ref{other}. In
section~\ref{connections} I discuss the possible relations between
magnetars and other classes of astrophysical objects. Some
prospects for future observations are given in the concluding
section.


\begin{landscape}

\begin{table}
\caption{Multiwavelength emission from AXPs and SGRs}
 \centering
\label{tab-list}
\begin{tabular}{lccccccl}
\hline\noalign{\smallskip}
Name$^{(a)}$   &Hard X-rays$^{(b)}$& Soft X-rays$^{(b)}$& Optical$^{(b)}$ &   IR$^{(b)}$  & Radio$^{(b)}$& Distance      & Location \\[3pt]
         &  ($>$10 keV)  & ($<$10 keV)   &       &      &    & (kpc)  &  \\[3pt]
\tableheadseprule\noalign{\smallskip}
\multicolumn{8}{c}{\textbf{Anomalous X--ray Pulsars}} \\ [3pt]
 \hline
\smc     & -       & P       &  -     & -       & -     & 61     & SMC \\[3pt]
\cite{lam02}  &    & \cite{lam02}  &      &         &       &        &     \\[3pt]
 \hline
\uu      & P       & P       & P       & D       & -     & 3.6    & \\[3pt]
\cite{mer95}      & \cite{den06}     & \cite{isr94}     & \cite{ker02}     & \cite{hul04,wan06}    &       &\cite{dur06a}&     \\[3pt]
\hline
\oo      & D       & P       & -       & D       & -     & 9      & \\[3pt]
\cite{mer95}      & \cite{ley08}    & \cite{sew86}    &         &\cite{wan02,isr02}  &       &\cite{dur06a}    &     \\[3pt]
\hline
\qui     & -       & P,T    &   -     & -       & P     & 9      & SNR G327.24--0.13 \\[3pt]
\cite{gel07}     &         & \cite{hal07b}    &         &         &\cite{cam07c}   &\cite{cam07c}&     \\[3pt]
\hline
\cxo     &     -   & P,T     & -       & -       & -     & 3.9    & Massive Star Cluster \\[3pt]
\cite{mun06}    &         &\cite{mun07,isr07b}  &         &         &       &\cite{kot07}   &  Westerlund 1     \\[3pt]
\hline
\rxs     & P       & P       & -       & D?       & -     & 3.8    & \\[3pt]
\cite{sug97}     & \cite{kui06,goe07b}  &\cite{sug97}     &         & \cite{dur06,tes07}   &       &\cite{dur06a}     &     \\[3pt]
\hline
\xte     & -       & P,T     & -       &   D     & P     & 3.1    & \\[3pt]
\cite{ibr04}     &         &\cite{ibr04,got04}  &         &\cite{isr04a}     &\cite{hal05b,cam06} &\cite{dur06a}     &     \\[3pt]
\hline
\kes     & P       & P       & -       & D?      & -     & 8.5      & SNR Kes 73\\[3pt]
\cite{vas97}     & \cite{kui04,mol04}    & \cite{vas97}    &         &\cite{tes07}     &       & \cite{lea07}     &     \\[3pt]
\hline
\axj $^{(c)}$    & -       & P,T     & -       & -       & -     &    8.5    & SNR G29.6+0.1 \\[3pt]
 \cite{tor98}    &         & \cite{tor98}    &         &    &  & \cite{tor98}  &     \\[3pt]
\hline
\ee      & -       & P       & -       & D       & -     & 7.5    & SNR CTB 109 \\[3pt]
\cite{mer95}      &         &\cite{fah81}     &         & \cite{hul01}    &       & \cite{dur06a}    &     \\[3pt]
\hline
 \multicolumn{8}{c}{\textbf{Soft Gamma-ray Repeaters}} \\ [3pt]
 \hline
\lmc     & -       & P       & -       & -       & -     & 55     & LMC, SNR N49 \\[3pt]
\cite{cli80}     &         & \cite{rot94}    &         &         &       &        &     \\[3pt]
\hline
\sedici  & -       & D,T     & -       & -       & -     & 11     & \\[3pt]
\cite{woo99c}     &         & \cite{woo99c}    &         &         &       & \cite{cor99}   &     \\[3pt]
\hline
\zerosei & D       & P       & -       & D       & -     & 15     & Massive Star Cluster \\[3pt]
\cite{lar86}    &\cite{mer05a}     &\cite{kou98}     &         & \cite{kos05,isr05} &       & \cite{cor04}   &     \\[3pt]
\hline
\zerozero& D       & P       & -       & D?      & -     & 15     & Massive Star Cluster \\[3pt]
\cite{maz79}     & \cite{goe06b}   &\cite{hur99e}     &         & \cite{tes07}    &       & \cite{vrb00}   &     \\[3pt]
\hline
 \hline
\end{tabular}

 \textbf{Notes:}

 $^{(a)}$ Here and throughout the whole paper I
use abbreviated names. See Table \ref{tab-positions} for the full
names of these sources.

$^{(b)}$ D = detection; P = pulsations detected; T = transient

$^{(c)}$ Candidate AXP (no $\pdot$ measurement)

\end{table}

\end{landscape}

\section{Spectral  Properties}
\label{sec:1}

\subsection{X-ray Luminosity}
\label{lum}

AXPs were discovered as relatively  bright (several
milliCrabs\footnote{1 mCrab $\sim$2$\times10^{-11}$ erg cm$^{-2}$
s$^{-1}$ in the 2-10 keV range.}), persistent X--ray sources, and
similar fluxes were later found in the X-ray counterparts of
galactic SGRs. Although the lack of optical identifications
hampered accurate distance  estimates for the individual objects,
it was clear from their collective properties (high X-ray
absorption and distribution in the Galactic plane) that these
objects had characteristic distances of at least a few kpc. Such
values, supported in some cases by the distance estimates of the
associated SNRs, implied typical luminosities in the range
10$^{34-36}$ erg s$^{-1}$, clearly larger than the rotational
energy loss inferred from their period and $\pdot$
values\footnote{With the reasonable assumption that these objects
are neutron stars (moment of inertia I$_{NS}$=10$^{45}$ g cm$^2$);
the fact that white dwarfs have much larger moments of inertia
(I$_{WD}$ \gtsima 10$^{4}$ I$_{NS}$) led to propose models based
on isolated white dwarfs, powered by rotational energy
\citep{pac90,uso94}.}.

\citet{dur06a} studied the optical reddening versus distance in
the fields of six AXPs  in order to infer distances from the
absorption measured in X-rays. This led, in a few cases, to
significantly revised distance estimates (e.g. 9$\pm$1.7 kpc wrt
$\sim$3 kpc for \oo ; 3.1$\pm$0.5 kpc wrt $\sim$10 kpc for \xte ).
If confirmed, this result implies  that the persistent
luminosities of AXPs ($<$10 keV) are all tightly clustered around
1.3$\times$10$^{35}$ erg s$^{-1}$. This is quite interesting since
in the magnetar model this luminosity is the expected saturation
value above which rapid cooling of the NS interior is
effective\footnote{Assuming the same luminosity for \qui\ (not
included in the above analysis) favors its location at $\sim$9
kpc, consistent with its radio dispersion measure \citep{cam07c},
rather than the closer distance of 4 kpc suggested by its possible
association with star forming regions in the Crux-Scutum spiral
arm \citep{gel07}.} \citep{tho96}.

Unfortunately the method of \citet{dur06a} cannot be used for the
SGRs, since they are too far and absorbed. Assuming that they have
the same luminosity derived for the AXPs, one obtains d$\sim$10
kpc and $\sim$8 kpc, for \zerozero\ and \zerosei\ respectively,
while their possible associations with star clusters (sect.
\ref{ass}) favor slightly larger distances. Also \lmc\ in the
Large Magellanic Cloud, with a well known distance implying a
luminosity of $\sim10^{36}$ erg s$^{-1}$ \citep{kul03}, does not
fit in this picture. Thus there is some evidence that the SGRs
might have a slightly higher luminosity than the persistent AXPs.

\subsection{X-ray spectra}
\label{spectra}

\begin{figure*}
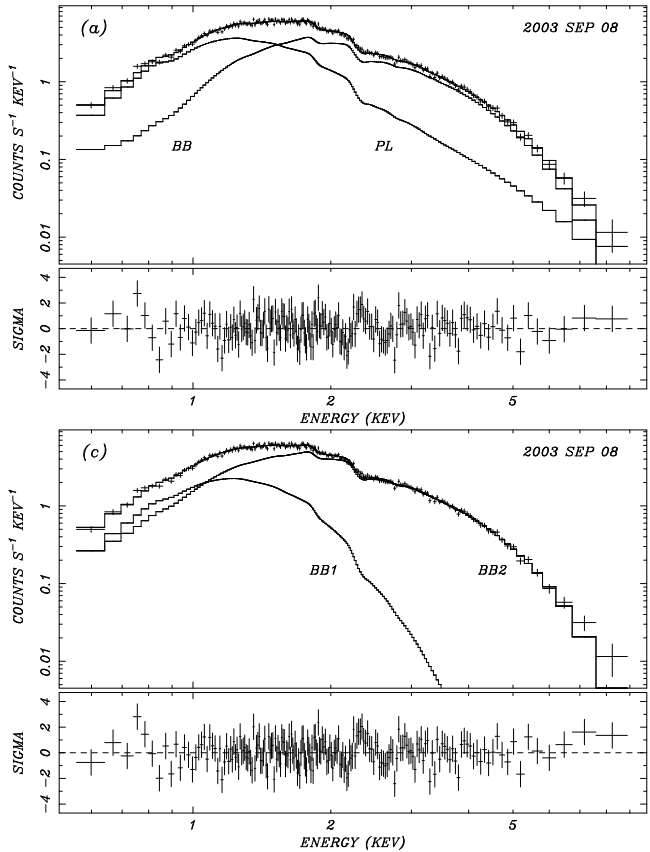

 \vbox{
\psfig{figure=j1810_spec_hal05a.eps,angle=-90,width=8.5cm}
\psfig{figure=j1810_spec_hal05b.eps,angle=-90,width=8.5cm}}
\caption{X--ray spectrum of \xte\ measured with the XMM-Newton
EPIC instrument (from \citet{hal05}). Equivalently good fits are
obtained with a power law plus  blackbody model (top panel) or
with the sum of two blackbodies (bottom panel). The second model
has the advantage that, when extrapolated to lower energies, it
does not exceed the optical and near infrared limits. }
\label{fig-j1810_spec}
\end{figure*}

AXPs have soft spectra below 10 keV, that are generally fitted by
a combination of a steep power-law (photon index $\sim$3-4) and a
blackbody with temperature kT$\sim$0.5 keV \citep{mer02}. In a few
AXPs, equivalently good fits are obtained with two blackbodies
(Fig.~\ref{fig-j1810_spec}) or other combinations of two spectral
components. Physical arguments in favor of the double-blackbody
spectrum were given by \citet{hal05}.

Although all these models are just phenomenological descriptions
of the data, they indicate that the soft X--ray emission is
predominantly of thermal origin, but the emerging spectrum is more
complex than a simple Planckian. This is not surprising,
considering the presence of a strongly magnetized atmosphere
and/or the effects of scattering in the magnetosphere. Several
attempts to correctly take into account these complex phenomena
have been done in recent years, leading to more physical spectral
models that seem promising to explain some of the observed
characteristics, such as the absence of cyclotron features and the
hard X--ray tails \citep{lyu06b,fer07,guv07,nob08}.

\begin{figure*}
\psfig{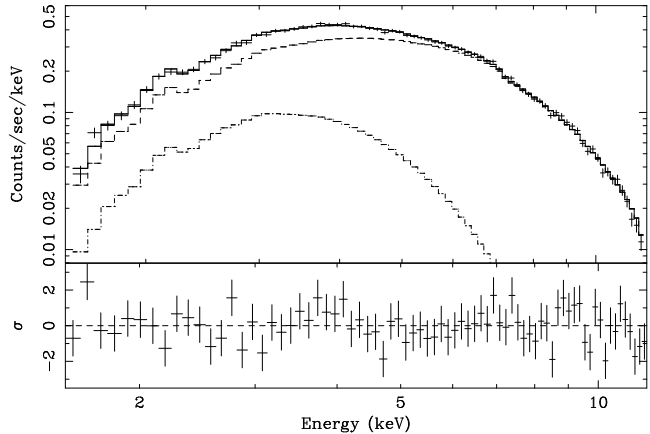}
\caption{XMM-Newton EPIC spectrum of \zerosei\ fitted with a power
law plus blackbody model (from \citet{mer05c}). The blackbody
component is the lower curve.  Notice, in comparison to
Fig.~\ref{fig-j1810_spec}, the much smaller relative contribution
of the blackbody component to the total flux.}
\label{fig-sgr1806_spec}
\end{figure*}

SGRs tend to have  harder spectra below 10 keV than AXPs, with the
exception of \lmc\, which is the most ``AXP-like'' of the SGRs.
They also suffer of a larger interstellar absorption, which makes
the detection of blackbody-like components more difficult. Most
spectra of \zerosei\ and \zerozero\ have been well fit with
power-laws of photon index $\sim$2. However, when good quality
spectra with adequate statistics are available, blackbody-like
components with kT$\sim$0.5 keV can be detected also in these
sources (Fig.~\ref{fig-sgr1806_spec})  \citep{mer05c,mer06b}.

\subsection{Features from cyclotron resonance scatter}
\label{lines}

In principle, a direct measurement of the neutron stars magnetic
field could come from the detection of spectral features due to
cyclotron resonance, provided that the particles (electrons or
ions) responsible for the effect are securely identified. While
electron lines would lie in the unobserved range above $\sim$1 MeV
for  magnetic field strengths of $\sim 10^{14}-10^{15}$~G, proton
cyclotron features are expected to lie in the X-ray range.

The first calculations of the spectrum emerging from the
atmospheres of magnetars in quiescence have confirmed this basic
expectation \citep{zan01,ho01}. Model spectra exhibit a strong
absorption line at the proton cyclotron resonance, $E_{c,p}\simeq
0.63 z_G(B/10^{14}\, {\rm G})$ keV, where $z_G$ is the
gravitational redshift, typically in the 0.70--0.85 range at the
neutron star surface. However, despite extensive searches no
convincingly significant lines have been detected up to now in the
persistent emission of magnetars\footnote{A report of a possible
feature in \rxs\ \citep{rea03}, has not been confirmed by better
data \citep{rea05a}.}. The tightest upper limits on the presence
of lines in the 1-10 keV range have been derived with XMM-Newton
 \citep{woo04,mer05c,tie05a,mer06b,rea07e} and Chandra \citep{jue02} observations.

Some reasons have been proposed to explain the absence of
cyclotron features, besides the obvious possibility that they lie
outside the sampled energy range. Magnetars  differ from ordinary
radio pulsars not only for the field strength, but also because,
as discussed in section \ref{twist}, their magnetospheres are
highly twisted and can support current flows \citep{tho02}. The
presence of charged particles (electrons and ions) produces a
large resonant scattering depth at frequencies depending on the
local value of the magnetic field, thus leading to the formation
of a hard tail instead of a narrow line.  A different explanation
for the lack of lines involves vacuum polarization effects. It has
been calculated that in strongly magnetized atmospheres this
effect can significantly reduce the equivalent width of cyclotron
lines, thus making  their detection more difficult \citep{ho03}.

The situation is possibly different for what concerns the bursts,
for which several line features have been reported in RXTE data,
although not always with high statistical significance. An
emission line at 6.4 keV was detected in \zerozero\ during the
precursor burst of the 1998 August 29 intermediate flare
\citep{str00}. Evidence for lines during some bursts  has also
been claimed for \zerosei\ \citep{ibr02,ibr03}. Lines were
reported also from AXPs: in two bursts from \oo\
\citep{gav02,gav06} and in single bursts from  \xte\ \citep{woo05}
and \uu\ \citep{gav08}. For these three sources the lines were at
$\sim$13--14  keV.

It has been proposed that such features are only visible in
bursts, and not in the quiescent emission, because during bursts
there is a higher photon flux and/or blown off baryons that
provide enough optical depth \citep{mer05c,rea05a}. However, due
to their sporadic appearance, and sometimes debatable statistical
significance, these   features require an independent
confirmation, possibly with a different instrument.

\subsection{Hard X-ray emission}
\label{hardx}

Until a few years ago, the detection of magnetars in the hard
X-ray range was limited to the bursts and flares from SGRs. The
discovery with the INTEGRAL satellite of persistent hard X-ray
tails extending to $\sim$150 keV in AXPs came as a surprise,
considering their soft spectra below 10 keV
\citep{kui04,mol04,den06}. The hardest spectra of SGRs made them
more promising targets for hard X-ray telescopes, and indeed some
indication for the presence of hard tails in SGRs were already
present in earlier data. For example, in 1997 BeppoSAX detected a
significant emission in the 20-150 keV range, most likely
originating from \zerozero\ \citep{esp07}. However, only with the
imaging capability of the INTEGRAL IBIS telescope  it was possible
to unambiguously confirm the presence of persistent hard X--ray
emission in two SGRs \citep{mer05a,goe06b}.

Emission above \gtsima20 keV has been detected for four AXPs and
two SGRs (see Table \ref{tab-list}). The upper limits on the
non-detected sources are not deep enough to exclude that they have
similar hard X-ray emission.  In most cases pulsations have also
been seen. Long term variability of the hard X--ray flux has been
significantly established for \zerosei\ \citep{mer05a} and \rxs\
\citep{goe07b}, and cannot be excluded in the other sources.

\begin{figure}
\psfig{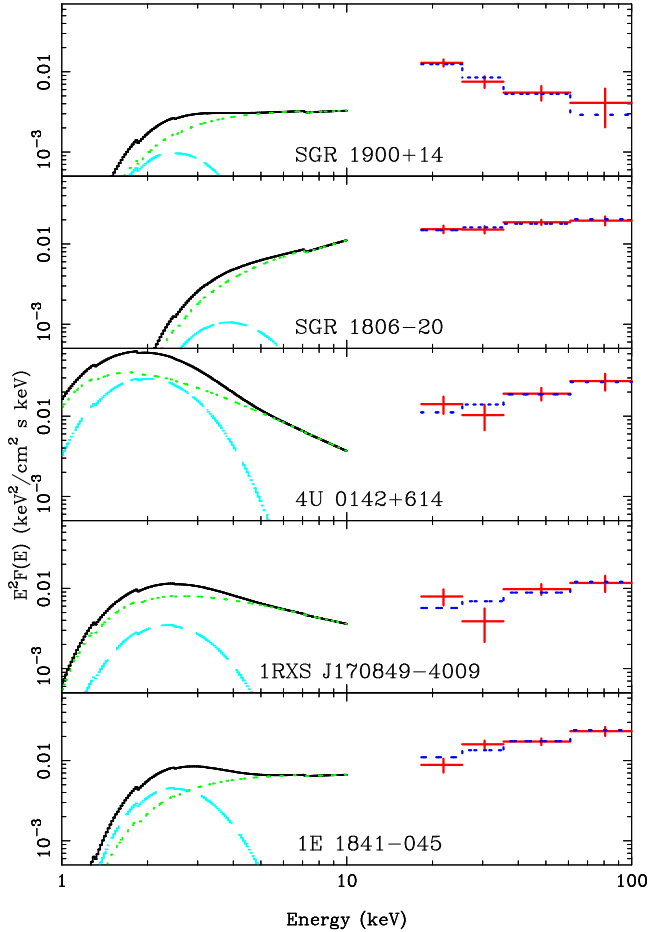}
\caption{XMM-Newton and INTEGRAL spectra of magnetars (from
\citet{goe06b}). Note the different behavior of SGRs (two top
panels) and AXPs: in the latter sources the spectra turn upward
above 10 keV, while in the SGRs the spectra steepen.}
\label{fig-integral_spec}
\end{figure}

\begin{figure}
 \vbox{
 \psfig{figure=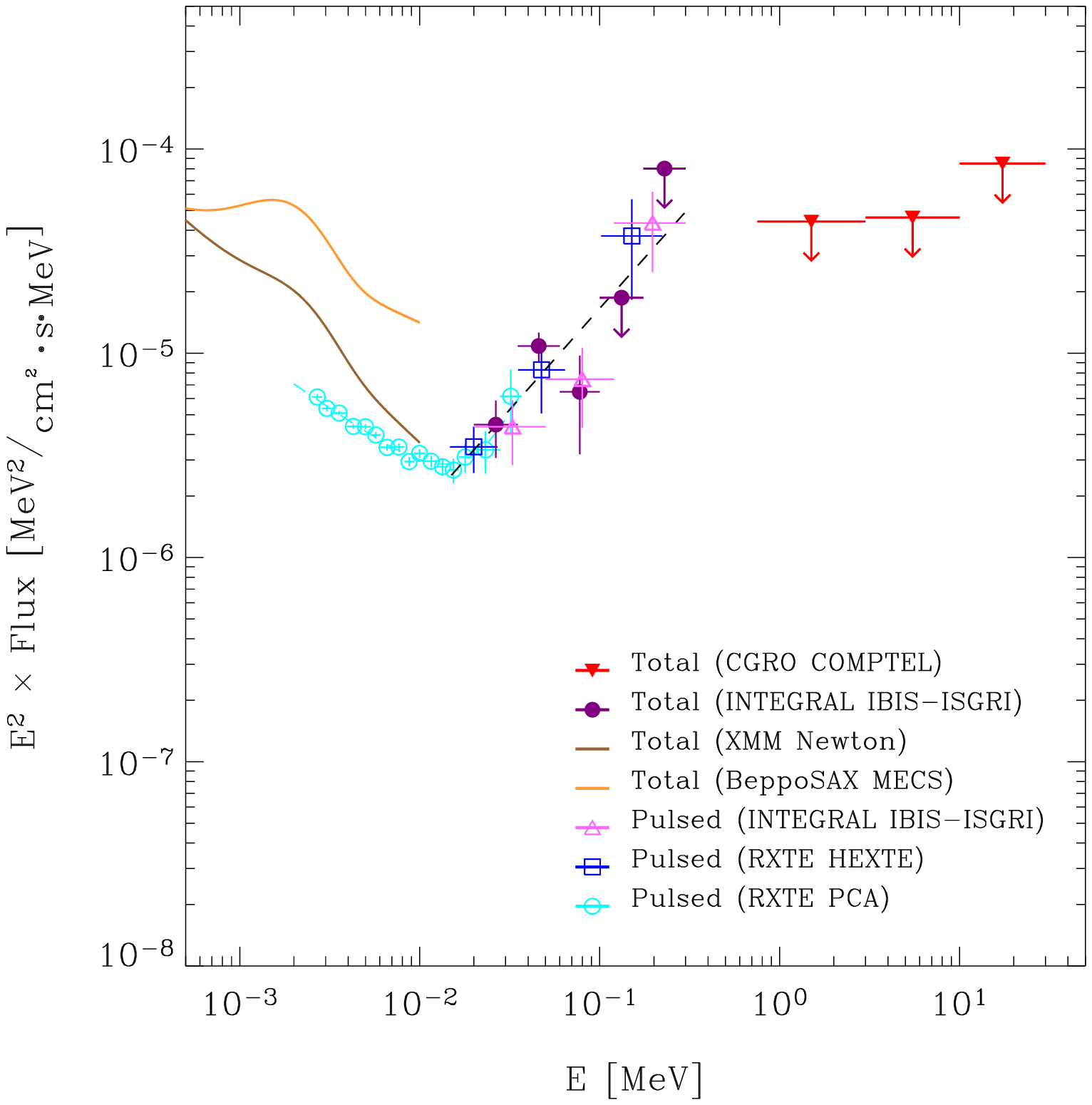,angle=0,width=8.cm}
 \psfig{figure=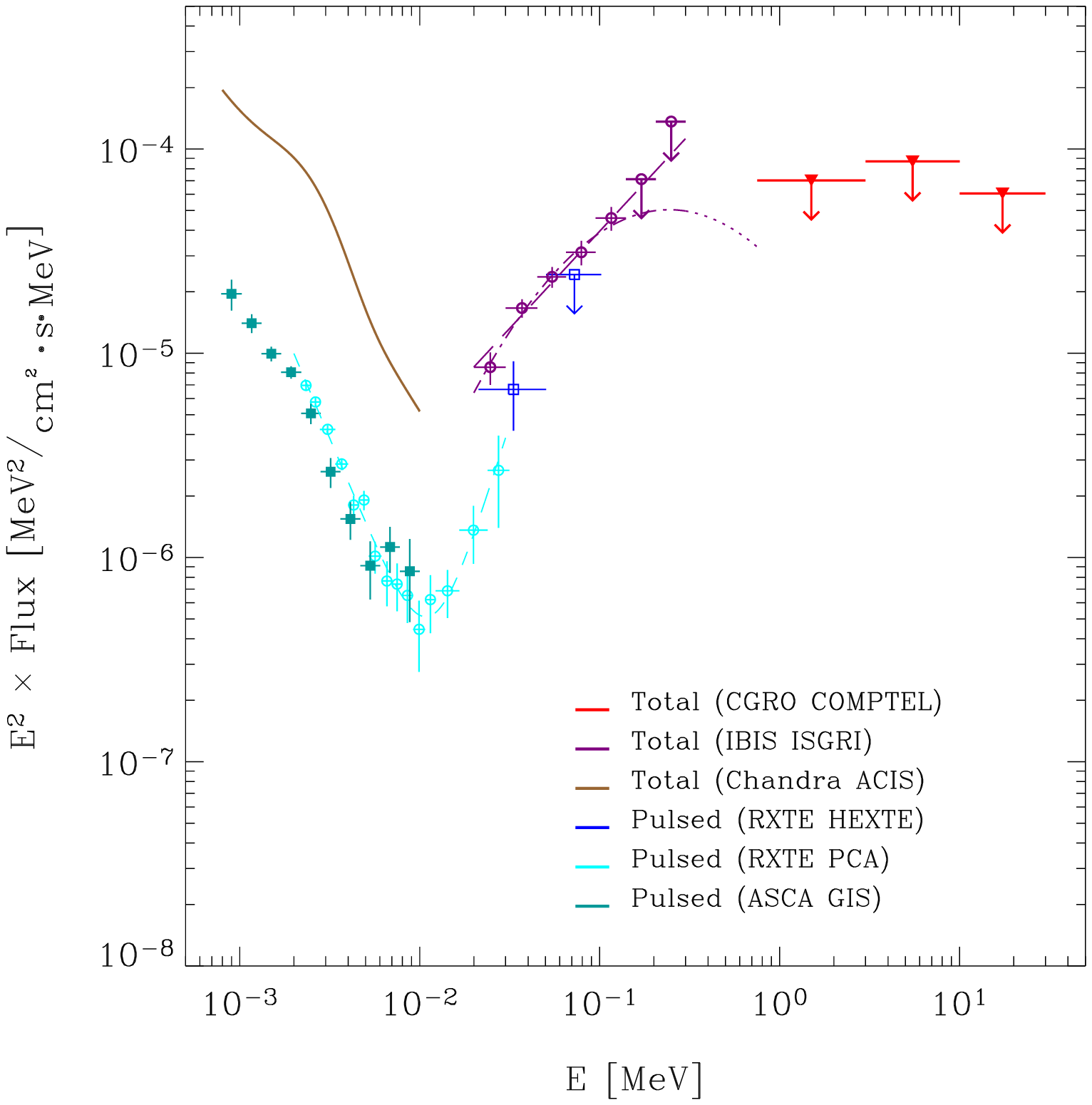,angle=0,width=8.cm}
 \psfig{figure=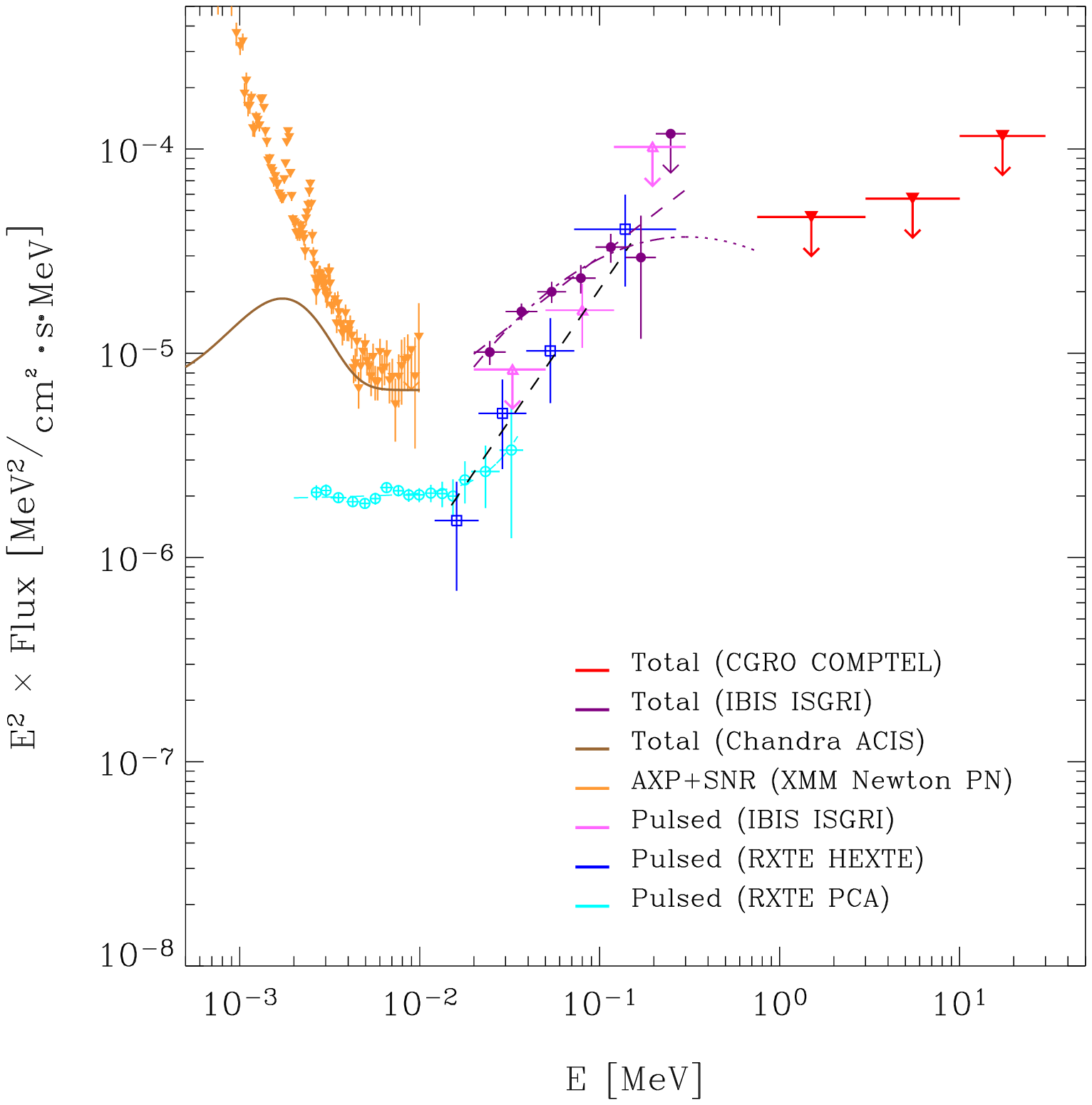,angle=0,width=8.cm}}
 \caption{Broad band spectra of Magnetars (from \citet{kui06}). From top to bottom:
 \rxs , \uu\ and \kes . Both the total and the pulsed emission are indicated.  }
\label{fig-xgamma_spec}
 \end{figure}

In the case of the AXPs,  the spectra above 20 keV are well fit
with rather hard power laws (Fig.~\ref{fig-xgamma_spec}), while
the spectra of SGRs are steeper (Fig.~\ref{fig-integral_spec}).
The power law photon indexes $\Gamma\sim$1--2 seen in the AXPs
\citep{kui06} imply a spectral flattening in the 10-20 keV range,
and indicate that the hard X--ray tails above 10 keV and the steep
power law often used in the spectral fits at lower energies are
two clearly distinct components. Most importantly, the flat
spectra imply that the energy released in the hard X-ray range is
a significant fraction of the total energy output from these
sources. The spectra obtained by considering only the pulsed flux
are harder than those of the total flux, indicating that the
pulsed fraction increases with energy. The most striking case is
\uu\ for which the pulsed emission has a power law photon index
$\Gamma=-0.8$ \citep{kui06}.

\section{Variability  Properties}
\label{variab}

\subsection{Short Bursts}
\label{burst}

\begin{figure}
\psfig{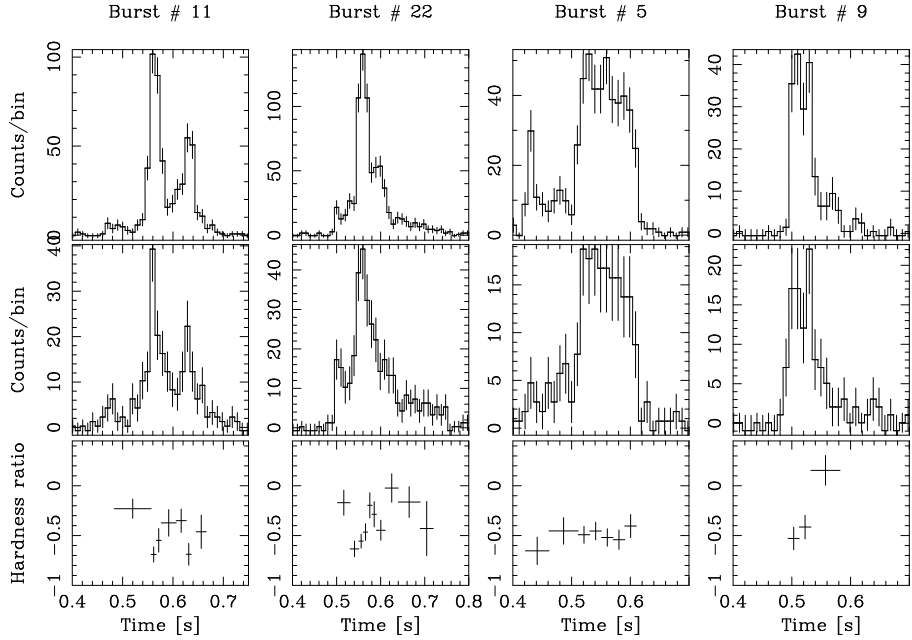}
\caption{Short bursts from \zerosei\ observed with the IBIS
instrument on board INTEGRAL (from \citet{goe04}). Top panels:
light curves in the soft energy range S=15-40 keV. Middle panels:
light curves in the hard energy range H=40-100 keV. Bottom panel:
hardness ratios, defined as (H-S)/(H+S),  showing that  spectral
evolution is present in some burst. } \label{fig-1806_bursts}
\end{figure}

SGRs are characterized by periods of activity during which they
emit numerous short bursts in the hard X-ray / soft gamma-ray
energy range. This is indeed the defining property that led to the
discovery of this class of high-energy sources. The bursts have
peak luminosity up to $\sim10^{42}$ erg s$^{-1}$ and durations
typically in the range $\sim$0.01-1 s, with a lognormal
distribution peaking at $\sim$0.1 s. Most of the bursts consist of
single or a few pulses with fast rise times, usually shorter than
the decay times. Some examples of bursts light curves are shown in
Fig.~\ref{fig-1806_bursts}.  The waiting time between bursts is
also distributed lognormally \citep{hur94} and no correlations
exist between the bursts intensity and waiting time. SGR bursts
occur randomly distributed in rotational phase.

The bursts observed fluences span the  range from a few 10$^{-10}$
to $\sim$10$^{-5}$ erg cm$^{-2}$, and follow a power law
distribution, with some evidence for a flattening at lower values
\citep{gog00,goe06}. Since the faintest end of the distribution
has been  explored with instruments operating at lower energy, it
is currently unclear whether the flattening reflects an energy or
an intensity dependence.

\begin{figure*}
 \vbox{
 \psfig{figure=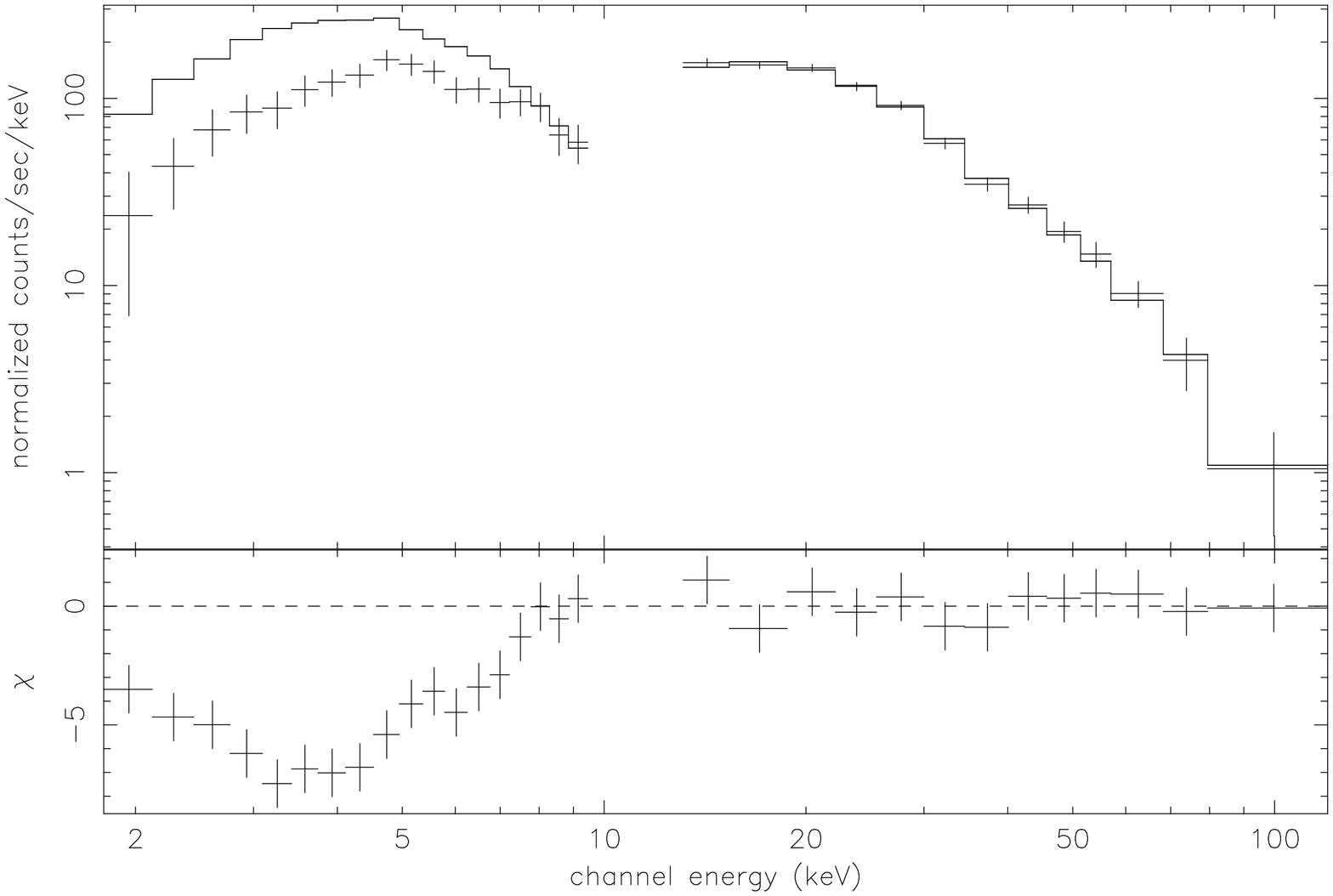,angle=-90,width=7cm}
 \psfig{figure=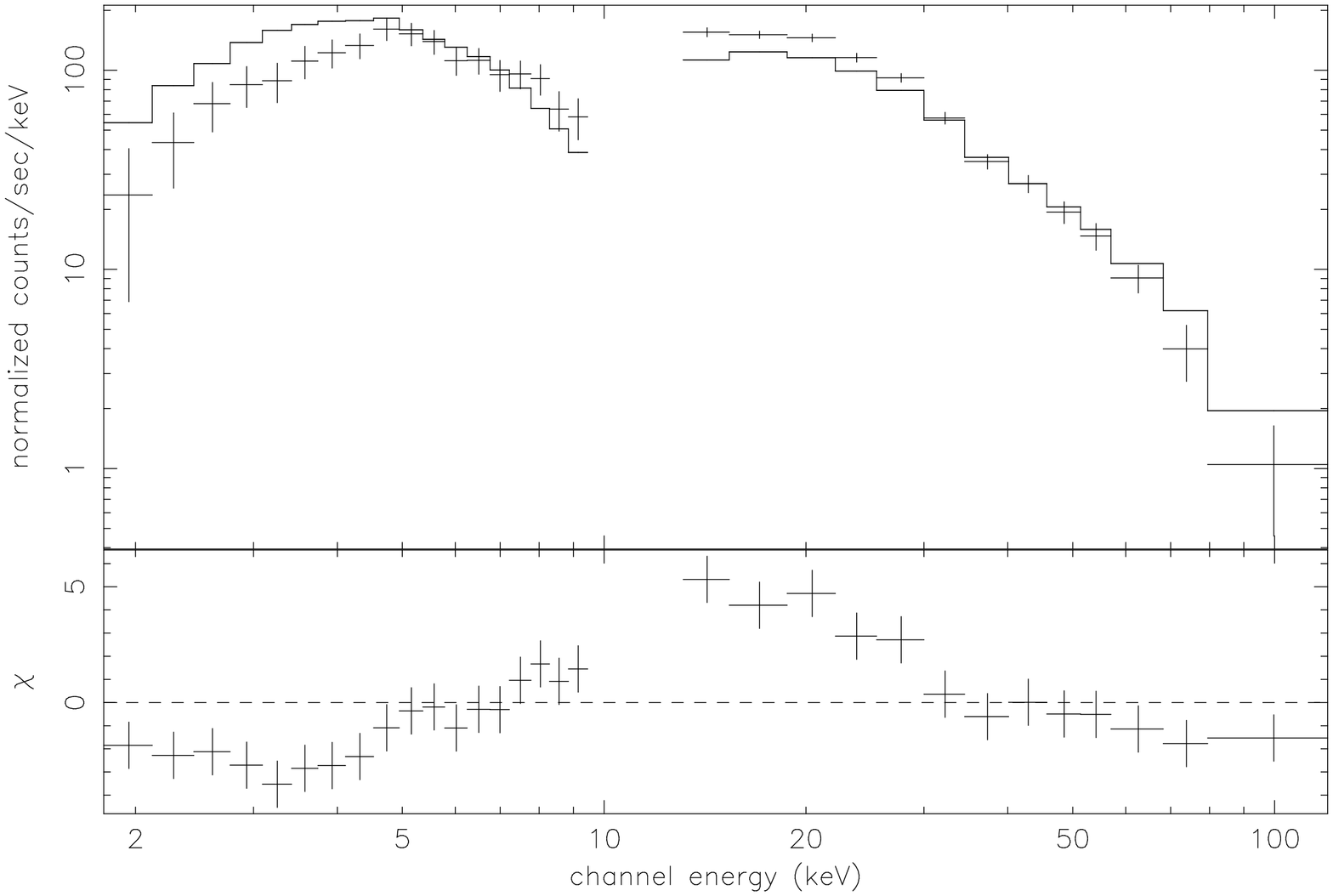,angle=-90,width=7cm}
 \psfig{figure=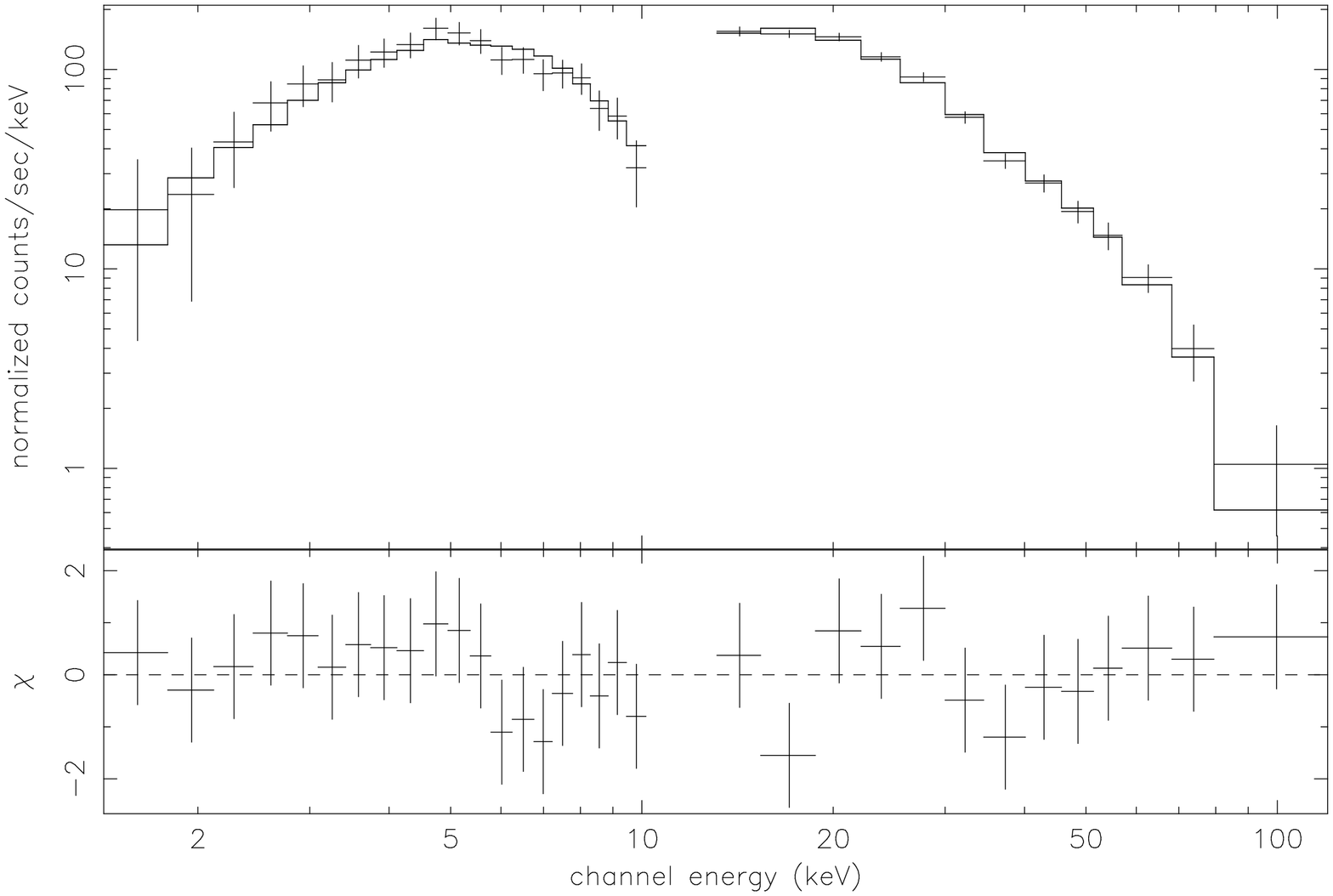,angle=-90,width=7cm}}
\caption{Cumulative spectrum of 10 bursts from \zerozero\ observed
with BeppoSAX (from \citet{fer04}). Data are from the MECS ($<$10
keV)  and PDS ($>$20 keV) instruments. \textit{Top panel:} an
optically thin thermal bremsstrahlung (OTTB) model is fitted only
to the PDS data; this model over predicts the flux in the MECS
energy range. \textit{Middle panel:} the OTTB model fitted to the
PDS and MECS data gives unacceptable residuals. \textit{Bottom
panel:} fit to the PDS and MECS data with the sum of two blackbody
models. }
\label{fig-burstspectra}
\end{figure*}

Until a few years ago, SGRs bursts were mainly observed above
$\sim$15 keV, where their spectra could be well fitted by
optically thin thermal\linebreak bremsstrahlung models with
kT$\sim$30--40 keV. More recent observations extending to lower
energy ($\sim$1-2 keV) showed that, if the same absorption is
assumed for the burst and the persistent emission,   the
bremsstrahlung fits overestimate the low energy flux the bursts
\citep{fen94} (see Fig.~\ref{fig-burstspectra}. One solution is to
invoke a higher absorption for the bursts, but there are no strong
physical arguments to support this. Alternatively, good fits over
the broad energy range from 1 to 100 keV can be obtained with the
sum of two blackbody models with temperatures kT$_1$$\sim$2-4 keV
and kT$_2$$\sim$8-12 keV \citep{fer04,oli04,esp07b}.

The discovery with RXTE that also AXPs can emit short burst
\citep{kas00,kas03}, similar to those of the SGRs, confirmed the
link between these two classes of objects and supported the
application of the magnetar model also to the AXPs. Bursts have
now been detected in several AXPs (see Table \ref{tab-timing}).
According to \citet{woo05} their properties suggest the existence
of two distinct  classes: \textit{type A} bursts with short and
symmetric profiles, and longer \textit{type B} bursts with
extended tails lasting tens to hundreds seconds. The latter have
thermal spectra, tend to occur at the phases of pulse maximum, and
have only been observed in AXPs\footnote{Although  long decaying
tails have been sometimes observed also in SGRs, they occurred
only after very bright bursts and with very small ratios between
the energy in the tail and that in the burst.}. Although type A
bursts are the ones typically observed in SGRs, at least one AXP
(\ee ) showed both types of bursts. This indicates that, even if
possibly originating from different mechanisms, these are not
mutually exclusive. \citet{woo05} suggested  that type A bursts
are caused by magnetic reconnections and type B ones by crustal
fractures.

\subsection{Giant flares}
\label{gf}

\begin{figure}
\psfig{figure=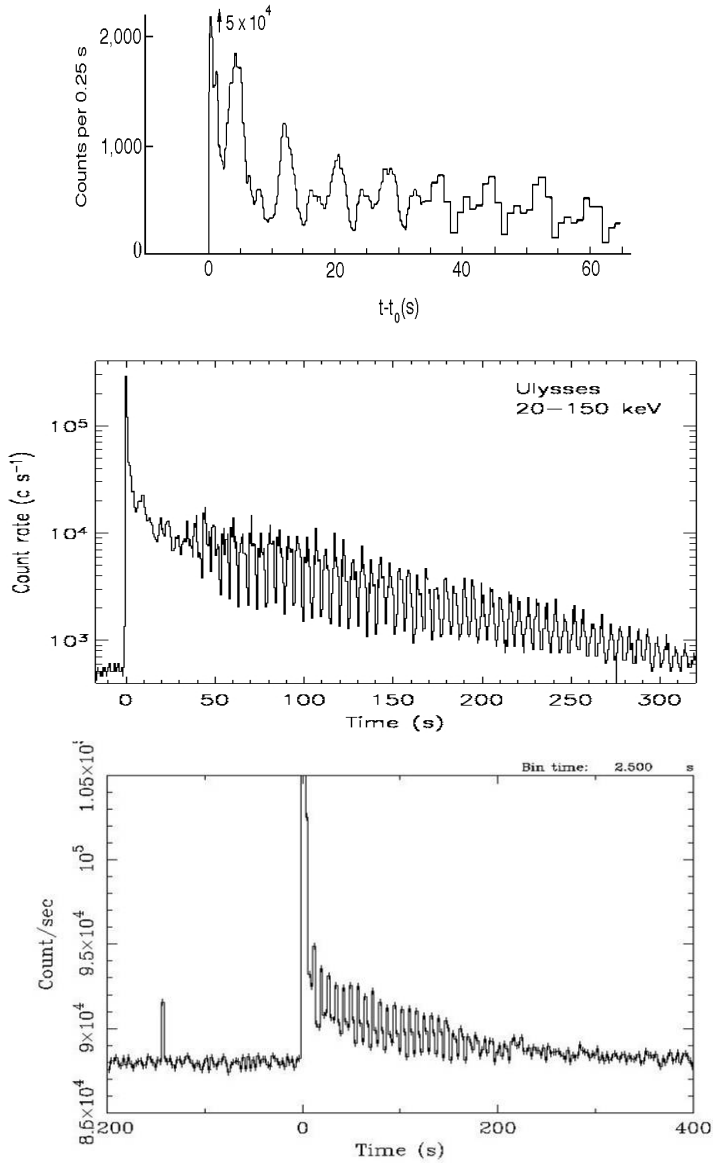,angle=0,width=13cm}
 \caption{Light
curves of the three giant flares from SGRs. Top panel: \lmc\
(Venera data in the 50-150 keV range, from \citet{maz82}); middle
panel: \zerozero\ (Ulysses data in the 20-150 keV range, courtesy
K. Hurley); bottom panel:  \zerosei\ (INTEGRAL SPI/ACS at E$>$80
keV, from \citet{mer05b}). The initial peaks of the flares for
\lmc\ and \zerosei\ are out of the vertical scale. }
\label{fig-GFLC}
\end{figure}

Giant flares have been observed so far only from SGRs (see Table
\ref{tab-gf}). They are characterized by the sudden release of an
enormous amount of energy ($\sim$(2--500)$\times10^{44}$ ergs), a
fraction of which escapes directly as a relativistically expanding
electron/positron plasma, while the remaining part is gradually
radiated by a thermal fireball trapped in the magnetosphere. This
gives to the giant flares a unique spectral and timing signature
consisting of a short hard spike followed by a longer pulsating
tail (Fig.~\ref{fig-GFLC}). These two characteristic
features\footnote{Other features that could be observed only in
some cases are a precursor and a long lasting afterglow.} have
been clearly recognized in the giant flares observed to date,
despite the differing quality and quantity of the available data.

The initial spikes of hard radiation reach a peak
luminosity\footnote{Here and in the following we quote
luminosities for isotropic emission.} larger than
$\sim4\times10^{44}$ erg s$^{-1}$ (up to a few 10$^{47}$ erg
s$^{-1}$ for \zerosei ). They are characterized by a rise time
smaller than a few milliseconds and a duration of a few tenths of
second.  Most detectors are saturated by the enormous photon flux
from these events. It is therefore particularly difficult to
reliably measure their peak fluxes and to reconstruct the true
shape of their light curves. Despite these difficulties, evidence
that the initial spikes have a complex, structured profile has
been reported for the 2004 giant flare of \zerosei\
\citep{ter05,sch05}.

It is well established that the spectra of the initial spikes,
with characteristic temperatures of hundreds of keV, are much
harder than those of the normal SGR short bursts. However, the
above caveats also apply to the spectral results, with the further
complication that, due to the long time intervals over which
spectra are generally accumulated, it is impossible to disentangle
the different time variable components. For example, for the
initial spike of \zerosei\ a cooling blackbody spectrum, with
temperature varying from 230 keV to 170 keV within $\sim$0.2 s,
was derived using charged particle detectors on the Wind and
RHESSI spacecrafts \citep{bog07}. Instead the analysis of the
radiation Compton-scattered from the Moon seen with the Coronas-F
satellite \citep{fre07}, as well as the results from small
particle detectors on other satellites \citet{pal05}, favor an
exponentially cut-off power-law, although with poorly constrained
parameters (photon index $\Gamma$=0.73$^{+0.47}_{-0.64}$ and
cut-off energy $E_o=666^{+1859}_{-368}$ keV).

The giant flares pulsating tails are characterized by a strong
evolution of the flux, timing and spectral properties. Their
spectra are softer than those of the initial spikes:  optically
thin bremsstrahlung models yield typical temperatures of a few
tens of keV. The better data available for the two more recent
giant flares required spectral models combining cooling thermal
components and power laws, sometimes extending into the MeV region
\citep{gui04,bog07,fre07}. The decaying light curves, observed for
a few minutes, are strongly modulated at the neutron star rotation
period, and show complex pulse profiles which evolve with time.

The energy emitted in the pulsating tails of the three giant
flares was roughly of the same order ($\sim10^{44}$ ergs), while
the energy in the initial spike of \zerosei\ (a few $10^{46}$
ergs) was at least two orders of magnitude higher than that of the
other giant flares (see Table~\ref{tab-gf}).  Since the tail
emission is thought to originate from the fraction of the energy
released in the initial spike that remains trapped in the neutron
star magnetosphere, forming an optically thick photon-pair plasma
\citep{tho95}, this indicates that the magnetic field in the three
sources is similar. In fact the amount of energy that can be
confined in this way is determined by the magnetic field strength,
which is thus inferred to be of several 10$^{14}$ G in these three
magnetars.

\begin{figure}
\psfig{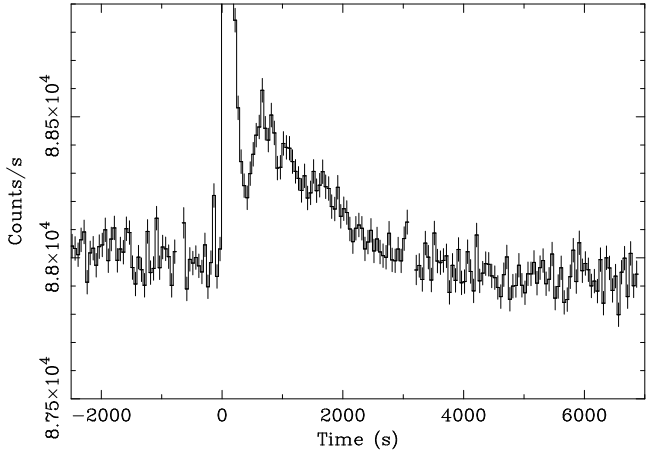}
\caption{SPI-ACS light curve of the  SGR 1806--20 giant flare
rebinned at 50 s to better show the emission lasting until one
hour after the start of the outburst (from \citet{mer05b}). Due to
this rebinning the pulsations at 7.56 s in the time interval 0-400
s cannot be seen in this plot.} \label{fig-1806GFaft}
\end{figure}

A unique  feature was detected in the \zerosei\ giant flare,
thanks to the large collecting area in the hard X-ray range ($>$80
keV) of the INTEGRAL/SPI Anti-Coincidence Shield (ACS). A hard
X-ray bump, peaking about 700 s after the start of the giant flare
and lasting about one hour was seen after the end of the pulsating
tail (Fig.~\ref{fig-1806GFaft}). Despite the lack of directional
information in the ACS and the non detection of pulsations, its
occurrence immediately after the giant flare strongly suggested to
associate this emission with \zerosei\ \citep{mer05b}. The reality
of this feature and its association with \zerosei\ has been
subsequently confirmed by independent detections, although with
smaller statistics and covering different time intervals, obtained
with Konus-Wind \citep{fre07} and RHESSI \citep{bog07} satellites.
The ACS data indicate a flux decay proportional to $\sim
t^{-0.85}$, and a fluence, in counts, similar to  that in the
pulsating tail (1-400 s time interval). Knowledge of the spectral
shape is required to convert the counts fluence into physical
units. The ACS does not provide any spectral resolution, but only
for hard spectra the ACS data can be reconciled with the small
fluence seen by RHESSI in the 3-200 keV range. This is also
consistent with the power law with photon index 1.6 derived from a
spectral analysis of the Konus-Wind data, which however refer to a
time interval after the INTEGRAL detection \citep{fre07}. Both the
power-law time decay and the hard power law spectrum suggest an
interpretation of this long-lasting hard X-ray emission in terms
of an afterglow, analog to the case of $\gamma$-ray bursts. In
fact, the presence of a relativistically expanding outflow
generated by the giant flare is also testified by the radio
observations of this event \citep{gae05,tay05,gra06}.

\begin{figure*}
 \vbox{
\psfig{figure=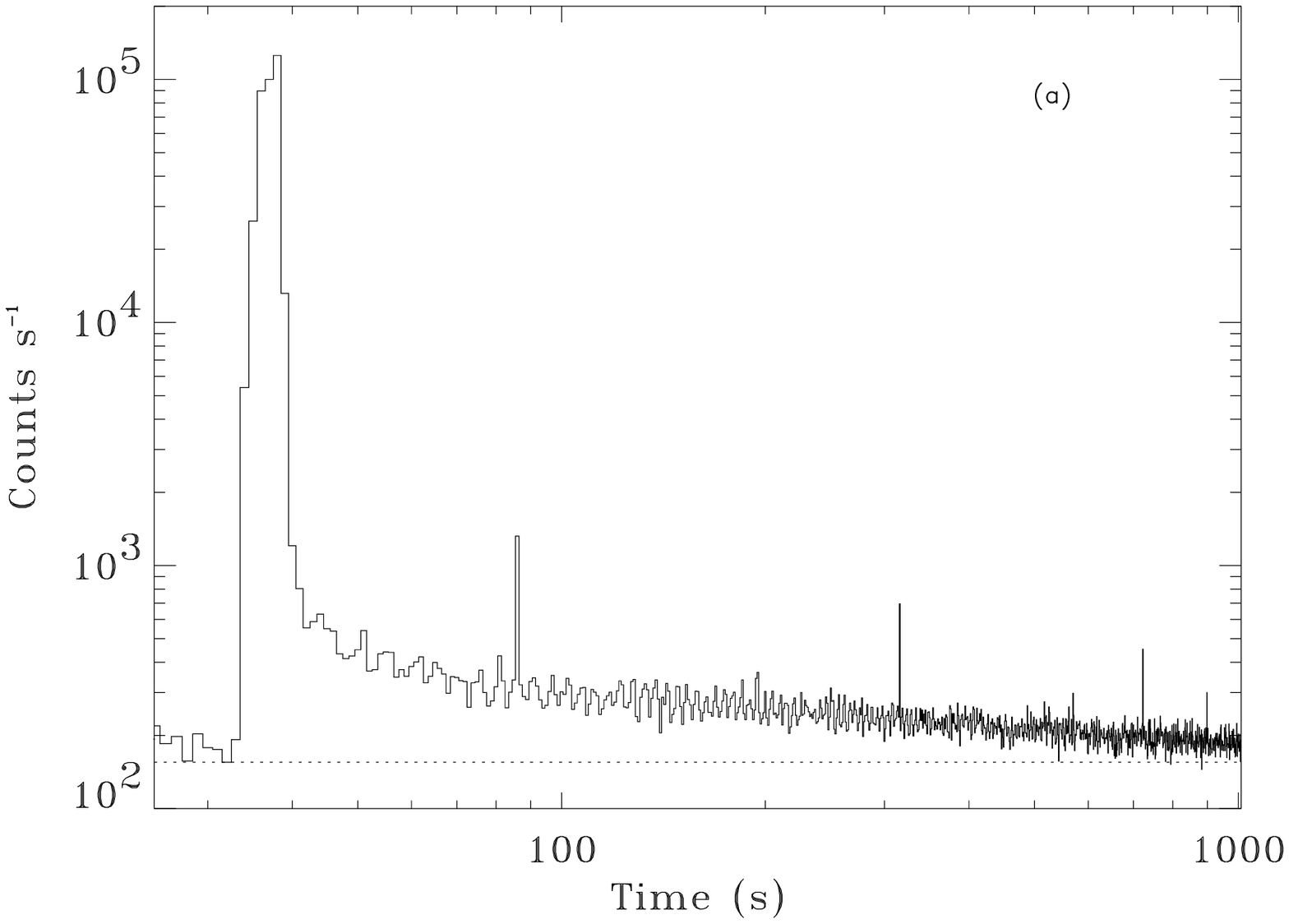,angle=0,width=8.5cm}
\psfig{figure=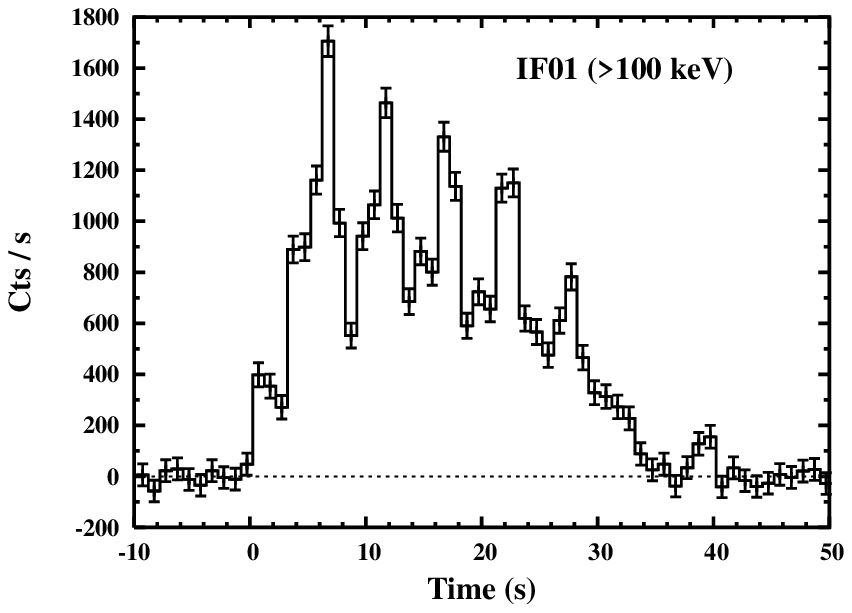,angle=0,width=8.5cm}}
\caption{Light curves of two intermediate flares from \zerozero .
Top panel: August 29, 1998 (RXTE, 2-90 keV, from \citet{ibr01}).
Bottom panel: April 18, 2001 (BeppoSAX GRBM, from \citet{gui04})}.
\label{fig-IFLC}
\end{figure*}

A few strong outbursts, involving a smaller energy than the giant
flares, but definitely brighter and much rarer than the normal
short bursts, have also  been seen in SGRs. They are therefore
called intermediate flares. The strongest one, lasting about 40 s,
was observed on April 18, 2001 from \zerozero\
\citep{kou01,gui04}. It was characterized by the presence of
pulsations at the neutron star rotation period, as in the tails of
giant flares, but without any initial spike (Fig.~\ref{fig-IFLC},
bottom panel). Other intermediate flares occurred in the same
source on August 29, 1998  \citep{ibr01},  only two days after the
giant flare, and on April 28, 2001 \citep{len03}.

\begin{table}[t]
\caption{Comparison of the three giant flares from SGRs}
\centering \label{tab-gf}
\begin{tabular}{cccc}
\hline\noalign{\smallskip}

 Source                      & \lmc\             & \zerozero\         & \zerosei\       \\ [3pt]
\tableheadseprule\noalign{\smallskip}

Date                 & March 5, 1979     &   August 27, 1998 & December 27, 2004 \\
Assumed distance            &  55 kpc           &  15 kpc &           15 kpc \\
 & & & \\
 \hline
\multicolumn{4}{c}{\textbf{Initial Spike}} \\ [3pt]
 \hline
Duration (s)                   & $\sim$0.25    & $\sim$0.35        & $\sim$0.5  \\
Peak luminosity (erg s$^{-1}$) & 3.6 10$^{44}$ & $>$8.3 10$^{44}$  & (2$\div5$) 10$^{47}$  \\
 Fluence (erg cm$^{-2}$)       & 4.5 10$^{-4}$ & $>$1.2 10$^{-2}$  & 0.6$\div$2  \\
Isotropic Energy (erg)         & 1.6 10$^{44}$ & $>$1.5 10$^{44}$  & (1.6$\div$5) 10$^{46}$ \\
  & & & \\
 \hline
\multicolumn{4}{c}{\textbf{Pulsating tail}} \\ [3pt]
 \hline
Duration (s)                  & $\sim$200      & $\sim$400         & $\sim$380 \\
 Fluence (erg cm$^{-2}$)      & 1 10$^{-3}$    & 9.4 10$^{-3}$     &  5 10$^{-3}$  \\
Isotropic Energy (erg)        & 3.6 10$^{44}$  & 1.2 10$^{44}$     & 1.3 10$^{44}$ \\
Spectrum                      & kT$\sim$30 keV & kT$\sim$20 keV    & kT$\sim$15--30 keV \\
Pulse Period (s)              & 8.1            & 5.15              & 7.56  \\
QPO Frequencies  (Hz)         & 43             & 28, 54, 84, 155   & 18, 30, 92.5, 150, \\
                              &                &                   & 625, 1480 \\
 \hline

\noalign{\smallskip}\hline
\end{tabular}
\end{table}

\subsection{Long term X-ray variability}
\label{variab}

\begin{figure*}
\psfig{figure=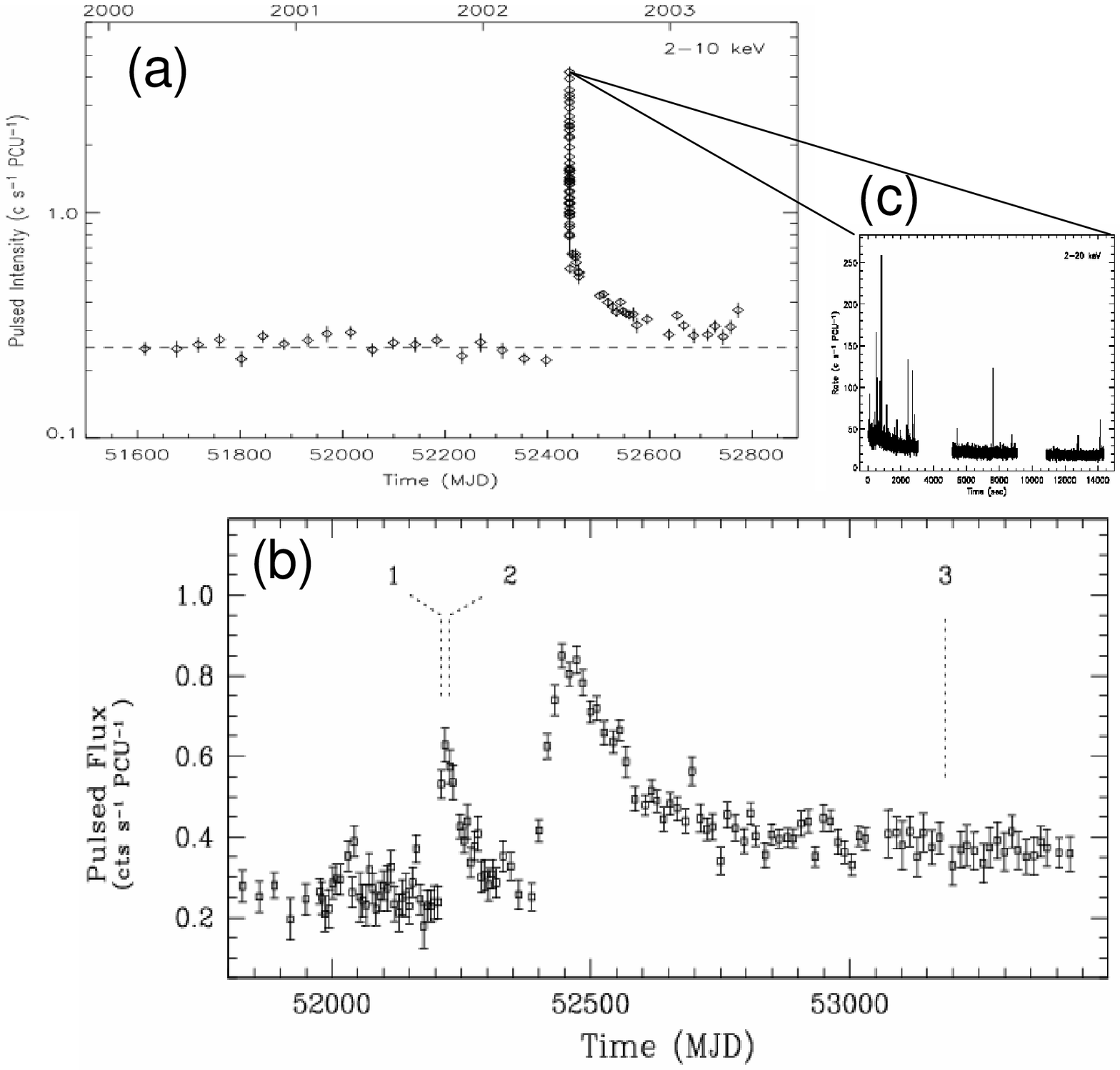,angle=0,width=12cm}
\caption{Comparison of the long term variability of the two AXPs
\ee\ (a) and \oo\ (b) (adapted from \citet{gav04} and
\citet{woo04}). All the panels show the \textit{pulsed} count rate
as measured with RXTE in the 2-10 keV range. Note that panels (a)
and (b) have approximately the same scale on the time axis, but
they differ in the vertical scale that is logarithmic for \ee .
The dashed lines in panel (b) indicate the times of the three
short bursts seen from \oo . Panel (c) gives an expanded view (1 s
bins) of the first four hours of the June 18, 2002 outburst from
\ee , during which many short bursts were detected}.
\label{fig-VAR_PULFLUX}
\end{figure*}

The apparent lack of pronounced variability, as strong as that
typical of accreting X-ray pulsars, was   among the distinctive
properties leading to the initial recognition of AXPs. Actually,
some indications for (small) long term variations were present in
early observations of some AXPs, but the fact that the data were
obtained with different satellites (some of which subject to
source confusion due to the lack of imaging capabilities) made
this evidence rather marginal. In the last decade, regular long
term monitoring has been provided by the RXTE satellite, but its
non-imaging instruments, unable to accurately estimate the
background for faint sources, have the drawback of precisely
measuring only the \textit{pulsed} component of the flux. Changes
in the pulsed flux might not reflect true luminosity variations if
the pulsed fraction is not constant, as well exemplified  by the
case of \oo\ discussed below. In the last years, especially thanks
to XMM-Newton and Chandra it has been possible to obtain much more
accurate flux measurements, and practically all magnetars for
which adequate data are available have shown some variability on
long timescales. Besides the most extreme cases of transients,
which span orders of magnitude in luminosity (section
\ref{trans}), at least two different kinds of long term variations
are present in magnetars. These are well demonstrated by the cases
of \oo\ and \ee discussed below and illustrated in
Fig.~\ref{fig-VAR_PULFLUX}.

For \oo , all the measurements obtained before July 2001 had
relatively large uncertainties and were  consistent\footnote{With
the possible exception of an upper limit implying a 10-fold lower
flux in December 1978 \citep{sew86}.} with an absorbed 2-10 keV
flux of $\sim5\times10^{-12}$ erg cm$^{-2}$ s$^{-1}$. Much better
data were subsequently obtained with XMM-Newton and Chandra,
showing unequivocal evidence for a large flux increase coupled to
a decrease in the pulsed fraction \citep{mer04}. The latter varied
from $\sim$91\%, when the source was at its ``historical''
luminosity level, to $\sim$55\% when the flux was more than two
times higher \citep{tie05a}. Continued monitoring with RXTE showed
that the high flux XMM-Newton and Chandra observations were
obtained during long lasting outbursts\footnote{I shall not use
the term ``flare'' often used to refer to these flux variations in
order to avoid confusion with the SGRs flares discussed in section
\ref{gf}.} in the \textit{pulsed} flux intensity \citep{gav04b}.
At the peak of the first outburst, which started in October 2001
and lasted about four months, short bursts were observed
\citep{gav02}. The second outburst, peaking in June 2002 was
brighter and much longer.

\begin{figure}
\psfig{figure=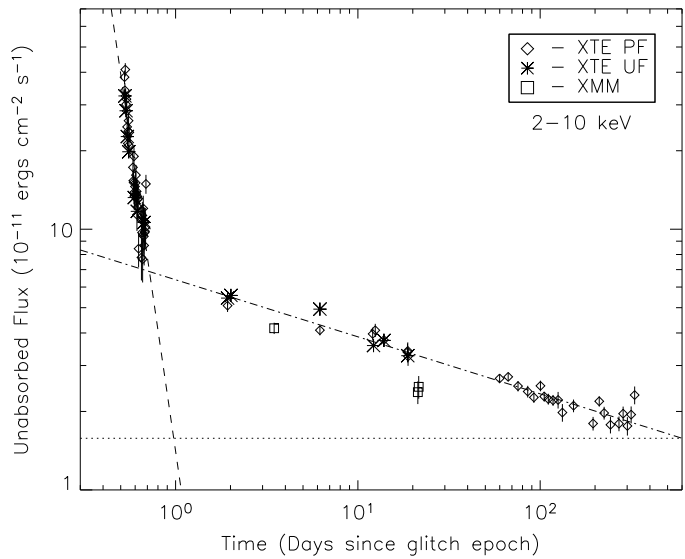,angle=0,width=10cm}
\caption{Long term flux evolution of \ee\  after the June 2002
outburst (from \citet{woo04}). During the first day the flux
evolution is well fit by a steep power law with temporal index
--4.5. At later times the much slower decay is well described by a
power law with index --0.2.} \label{fig-1e2259_lcurve}
\end{figure}

A different behavior was seen in \ee , when, in June 2002, RXTE
observed an outburst lasting a few hours during which many tens of
short bursts were emitted while  the pulsed and persistent X-ray
fluxes were more than one order of magnitude higher than in the
usual state \citep{kas03,woo04}. A large glitch was also observed
(see sect.~\ref{glit}). The initial rapid flux decay, accompanied
by significant evolution in the spectrum and pulse profile, was
followed by a slower decline lasting months (see
Fig.~\ref{fig-1e2259_lcurve}).

These two examples show that long term variations can occur either
as gradual changes in the flux, often accompanied by variations in
the spectrum, pulse profiles, and spin-down rate, or as sudden
outbursts associated with energetic events occurring on short
timescales, such as glitches and bursts\footnote{These short
bursting episodes can easily be missed in sparse observations of
magnetars; indeed some variability had already been reported in
\ee\ when two GINGA observations spaced by six months showed a
factor two luminosity increase coupled with a significant change
in the pulse profile \citep{iwa92}.}. In the first case it is
possible that the variations are driven by plastic deformations in
the crust causing changes in the magnetic currents configurations.
As discussed below (sect.~\ref{twist}), the currents supported in
twisted magnetospheres are ultimately responsible for the X--ray
emission through resonant cyclotron scattering and surface
heating. The more violent outbursts related to glitches and
bursting activity could instead be due to sudden reconfigurations
of the magnetosphere, when unstable conditions are reached. This
can probably occur on a large range of involved energies, with the
most extreme cases being the giant flares of SGRs
(sect.~\ref{gf}). The subsequent cooling of the neutron star
crust, heated in these events, can give rise to the observed long
term decays in the soft X-ray emission.

In March 2007, \oo\ showed an outburst \citep{tam07} similar to
the June 2002 event of \ee . The 2-10 keV flux measured with Swift
and Chandra soon after this event was the highest ever seen from
\oo , a factor 7 larger than the historical level. This event
demonstrates that the two kinds of variability are not mutually
exclusive and can occur in the same source.

\subsection{Transients}
\label{trans}

Transient X-ray sources have always been of great interest since
they allow to explore the theoretical models over a large
luminosity range and with fixed source parameters such as
distance, orientation, and, presumably, magnetic field. Some
evidence for the existence of transient magnetars came first from
the serendipitous observation in  December 1993 of \axj , a 7 s
pulsar with some characteristics of AXPs \citep{tor98} located in
the supernova remnant G29.6+0.1 \citep{gae99}. All the subsequent
observations of its error region detected only much fainter
sources \citep{vas00,tam06} suggesting the interesting possibility
of a transient, but failing to confirm the AXP nature of this
source by measuring a spin-down.

The discovery of \xte\  provided a much stronger case confirming
the existence of transient AXPs. Its outburst started before 2003
January 23, when the source was discovered with RXTE \citep{ibr04}
at a flux of $\sim6\times10^{-11}$ erg cm$^{-2}$ s$^{-1}$,   a
factor 100 higher than that of its quiescent counterpart recovered
\textit{a posteriori} in archival data. Since January 2003 its
luminosity decreased monotonically and is now approaching the
pre-outburst level (Fig.~\ref{fig-xtej1810_lcurve}). During the
outburst the spectral and timing properties of \xte\ were similar
to those of the persistent AXPs, and short burst were also
observed \citep{woo05}. X-ray observations carried out during its
long outburst decay showed a significant evolution of the spectrum
and pulse profile \citep{hal05,got05}. \citet{got07b}  found that
the spectrum is well described by two blackbody components whose
luminosity decreases exponentially with different timescales. The
temperature of the cooler component, initially at kT$_1$$\sim$0.25
keV, has been steadily decreasing since mid 2004, while at the
same time its emitting area expanded to cover almost the whole
neutron star surface. The hotter component cooled from
kT$_2\sim$0.7 keV to kT$_2\sim$0.45 keV while its emitting area,
initially $\sim$30 km$^2$, reduced by a factor $\sim$8. This
behavior has been interpreted in the framework of the magnetar
coronal model \citep{bel07} attributing the high temperature
component to a hot spot at the footprint of an active magnetic
loop and the cooler component to deep crustal heating in a large
fraction of the star. As discussed below, this object is also the
first magnetar from which pulsed radio emission has been detected
\citep{cam06}.

Two other transient AXPs have been identified recently: \linebreak
\cxo , in the young star cluster Westerlund 1 \citep{mun06}, and
\qui , likely associated to a possible SNR \citep{gel07}. The
first one spanned a dynamical range in luminosity larger than a
factor $\sim$300. \qui\ was seen to vary only by a factor $\sim$16
\citep{hal07b}, but it is possible that the peak of the outburst
was missed. Its pulsed radio emission makes it similar to the
prototype AXP transient \xte .

Only one of the four confirmed SGRs showed a transient
behavior:\linebreak \sedici\ was discovered in 1998, when more
than 100 bursts in about six weeks were observed with different
satellites \citep{woo99c}. No other bursts have been reported
since then. Its soft X-ray counterpart was identified with
BeppoSAX in 1998 at a luminosity level of \mbox{$\sim 10^{35}$ erg
s$^{-1}$}. Observations carried out in the following seven years
showed a monotonic decrease in its luminosity,  down to a level of
$\sim 4\times 10^{33}$ erg\, s$^{-1}$
(Fig.~\ref{fig-sgr1627_lcurve}). The latest XMM-Newton and Chandra
observations suggest that the flux stabilized at a steady level,
but they are affected by relatively large uncertainties and are
also compatible with a further gradual decay \citep{mer06a}.

The behavior of \sedici\ suggests a connection between the
bursting activity and the luminosity of transient magnetars. The
source  high state coincided with a period of strong bursting
activity, while in the following years, during which no bursts
were emitted,  its luminosity decreased. On the other hand,
\zerosei\ and \zerozero\ alternated periods with and without
bursts emission, but their X--ray luminosity did not vary by more
than a factor two. Even more remarkably, \lmc\  has a high
luminosity, despite being burst-inactive since 1979. This is
actually the most luminous of the SGRs, although its spectrum is
rather soft and similar to those of the AXPs.

The existence of transient magnetars,  with  quiescent
luminosities so small to prevent their discovery and/or
classification, has also implications for the total number of
magnetars in the Galaxy and their inferred birthrate. It is in
fact likely that there is a large number of undiscovered magnetars
currently in a low luminosity, quiescent state.

\begin{figure}
\psfig{figure=j1810_flux_got07b.ps,angle=-90,width=8.5cm}
\caption{X-ray light curve of the outburst of the transient AXP
\xte\ (from \citet{got07b})} \label{fig-xtej1810_lcurve}
\end{figure}

\begin{figure}
\psfig{figure=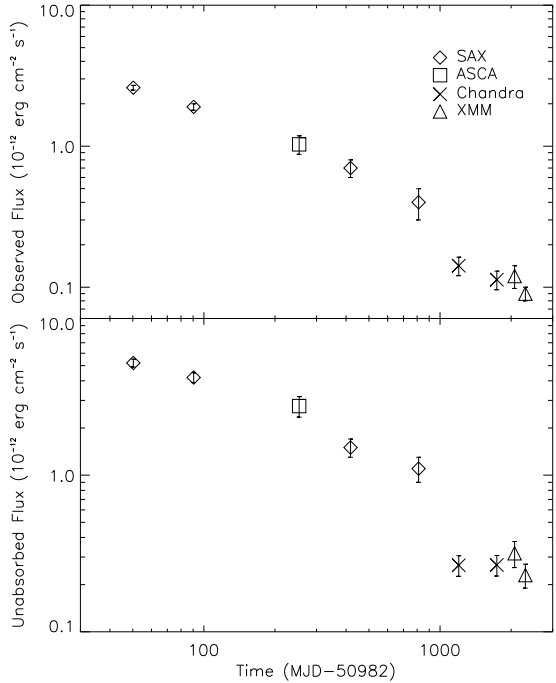,angle=0,width=8cm} \caption{Long term
flux decay of \sedici\ (from \citet{mer06a}). Note the differences
between the shape of the decays of the observed and unabsorbed
fluxes (2-10 keV). The latter are subject to large uncertainties,
especially at low fluxes, due to the poorly constrained spectra. }
\label{fig-sgr1627_lcurve}
\end{figure}

\section{Timing Properties}
\label{timing}

\subsection{Periods and period evolution}
\label{periods}

The narrow distribution  of spin periods  was among the
characterizing properties that led to recognize the AXP as a
separate class of objects \citep{mer95}. The period range of the
initial AXP group  (6--12 s) has long remained unchanged with an
almost tripled sample, and only recently it has been slightly
extended with the discovery of an AXP,  \qui , with a spin period
of 2.1 s \citep{cam07c}. This is still an extremely narrow
distribution, compared to that of X-ray binaries  (from
milliseconds to hours) and radio pulsars (from 1.4 ms to 8.5 s).

While the lack of observed magnetars with periods smaller than a
few seconds is easily explained by an early phase of rapid
spin-down, the absence of slowly rotating objects requires some
explanation \citep{psa02}, which, independently on the details,
imply that their lifetime as bright X--ray sources is limited. In
the magnetar models this could be caused, e.g. by the decay of the
magnetic field \citep{col00}.

The presence of periodic pulsations played an important role in
the early recognition of AXPs also because it allowed to search
for orbital Doppler modulations. Deep searches with RXTE failed to
see any signatures of orbital motion and allowed to set stringent
upper limits on the masses of potential companion stars
\citep{mer98,wil99}, showing that these objects were fundamentally
different from the high mass X-ray binary pulsars.

Long term variations in the spin-down rate were already evident in
some of the early AXP observations \citep{mer95b,bay96}, and were
later studied in great detail thanks to phase connected timing
analysis with the RXTE satellite \citep{kas99,kas01,gav02b,woo02}.
These observations indicate that the magnetars have a level of
timing noise larger than that typically observed in radio pulsars.
The timing noise is larger in the SGRs.  The presence of large
variations in the spin-down rate occurring on short timescales has
also been confirmed by accurate timing of the radio pulses in
\xte\ \citep{cam07b}. In addition to these gradual changes also
glitches have been observed in several magnetars (section
\ref{glit}).

An overall correlation between spin-down rate and spectral
hardness, with the SGRs showing the hardest spectra and larger
$\pdot$, was found by \citet{mar01}. This correlation is broadly
followed also in the long term variations  of the same source
\citep{mer05c}, and finds a natural explanation in the twisted
magnetosphere model (sect.~\ref{twist}). However, data with a more
continuous temporal coverage indicate that the situation is
actually more complex, with some of the $\pdot$ variations not
strictly correlated to large spectral or flux changes
\cite{kas01,woo06}.

\begin{figure*}
 \vbox{
 \hbox{
 \psfig{figure=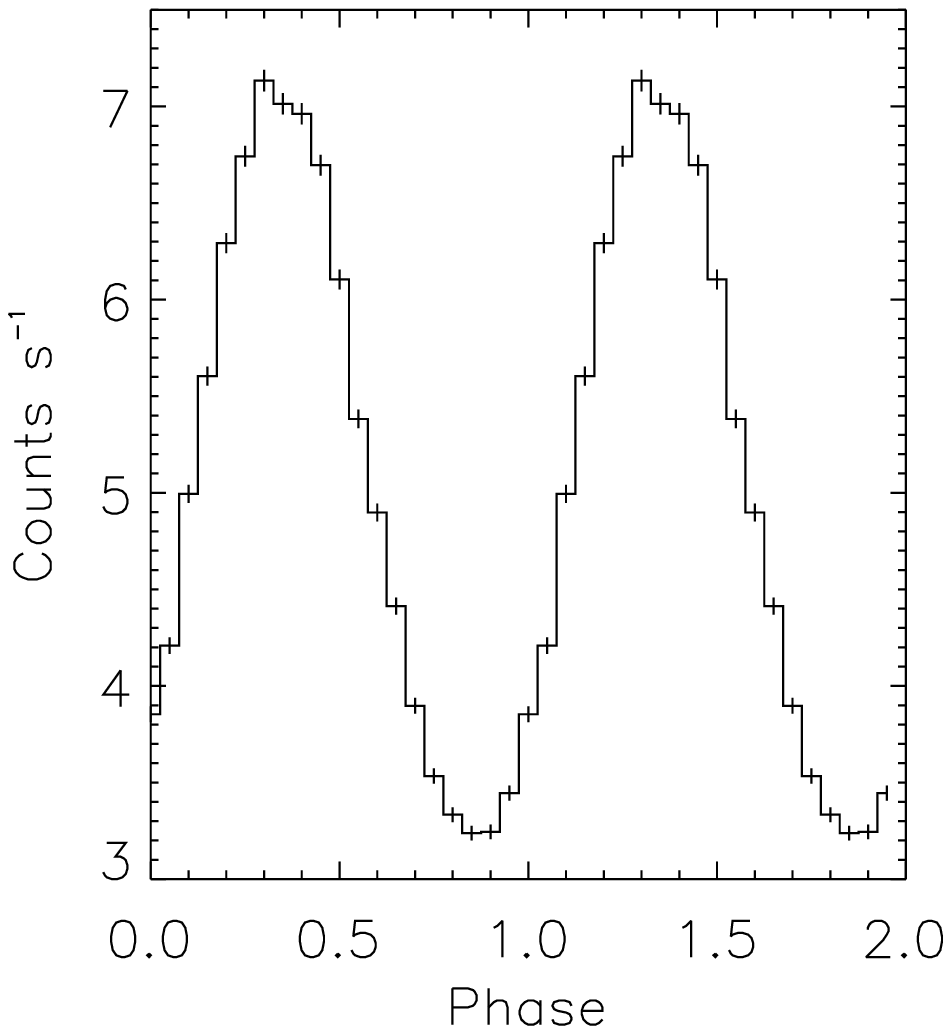,angle=0,width=5cm}
 \psfig{figure=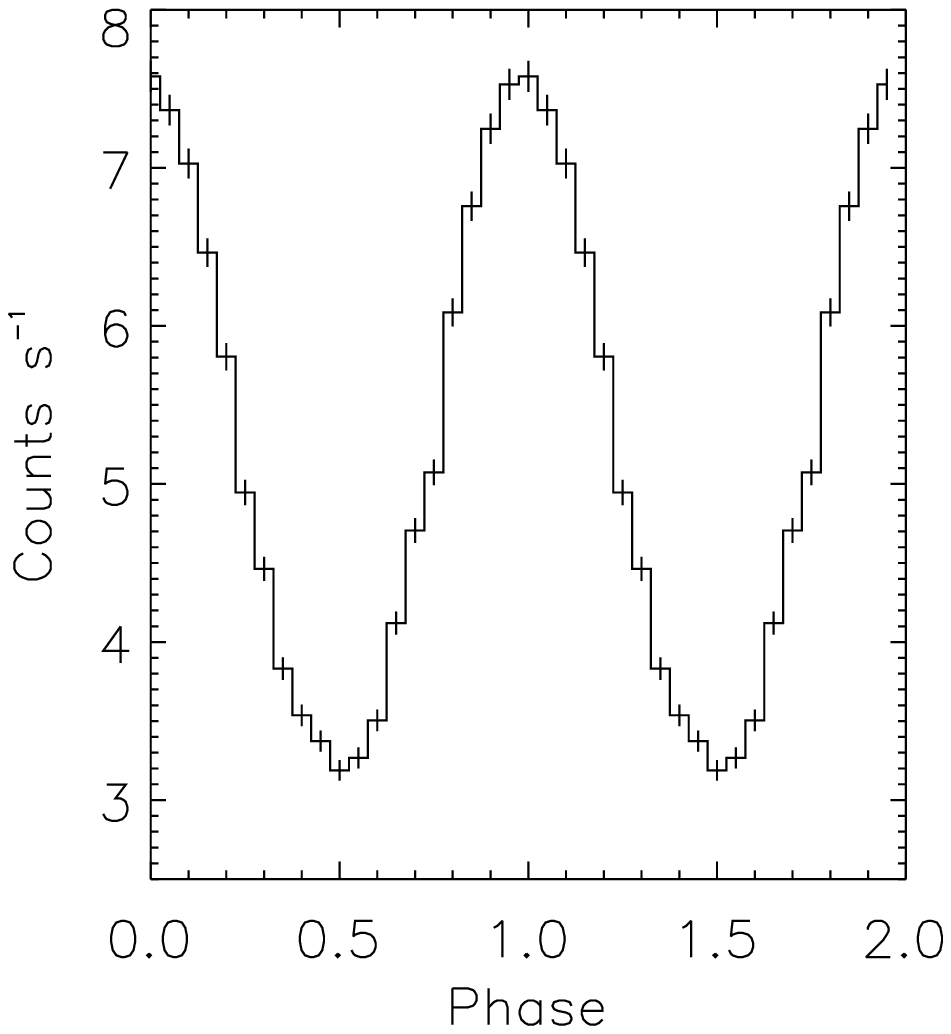,angle=0,width=5cm}
 }
\hbox{
 \psfig{figure=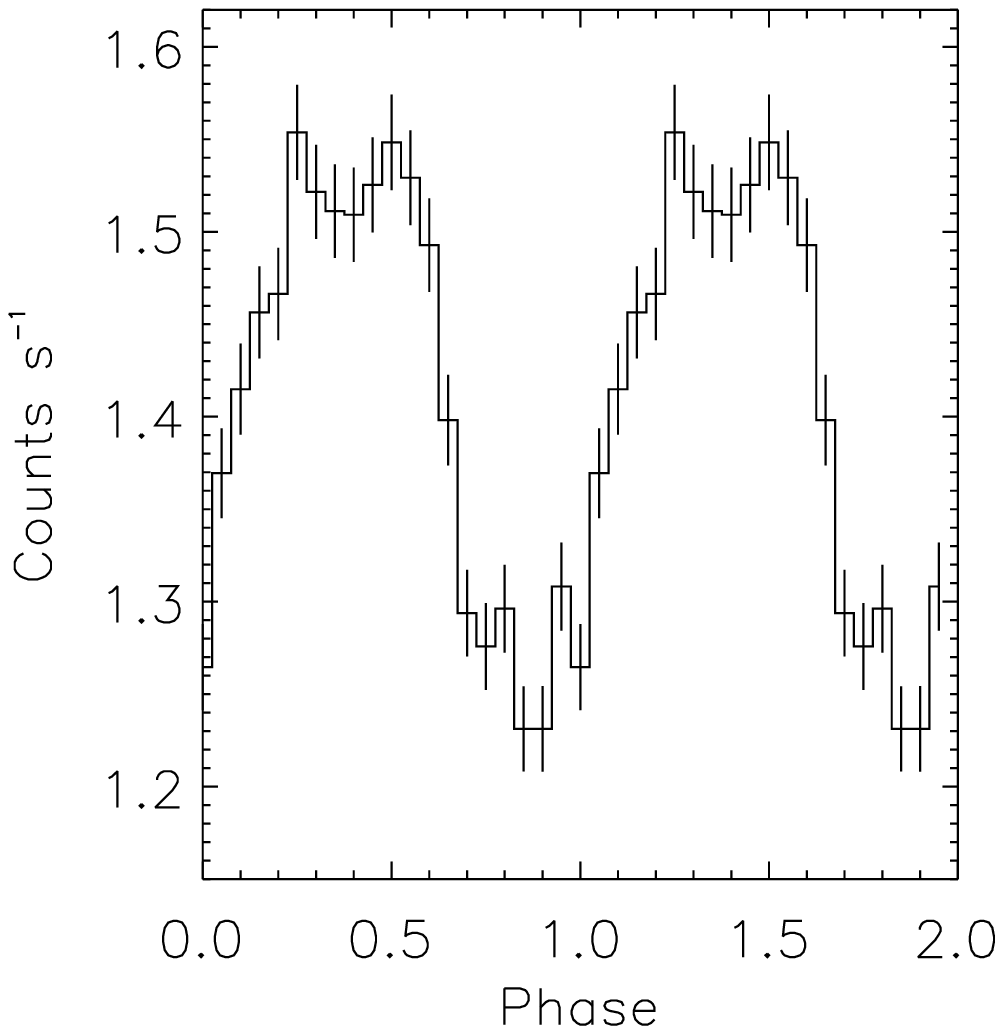,angle=0,width=5cm}
 \psfig{figure=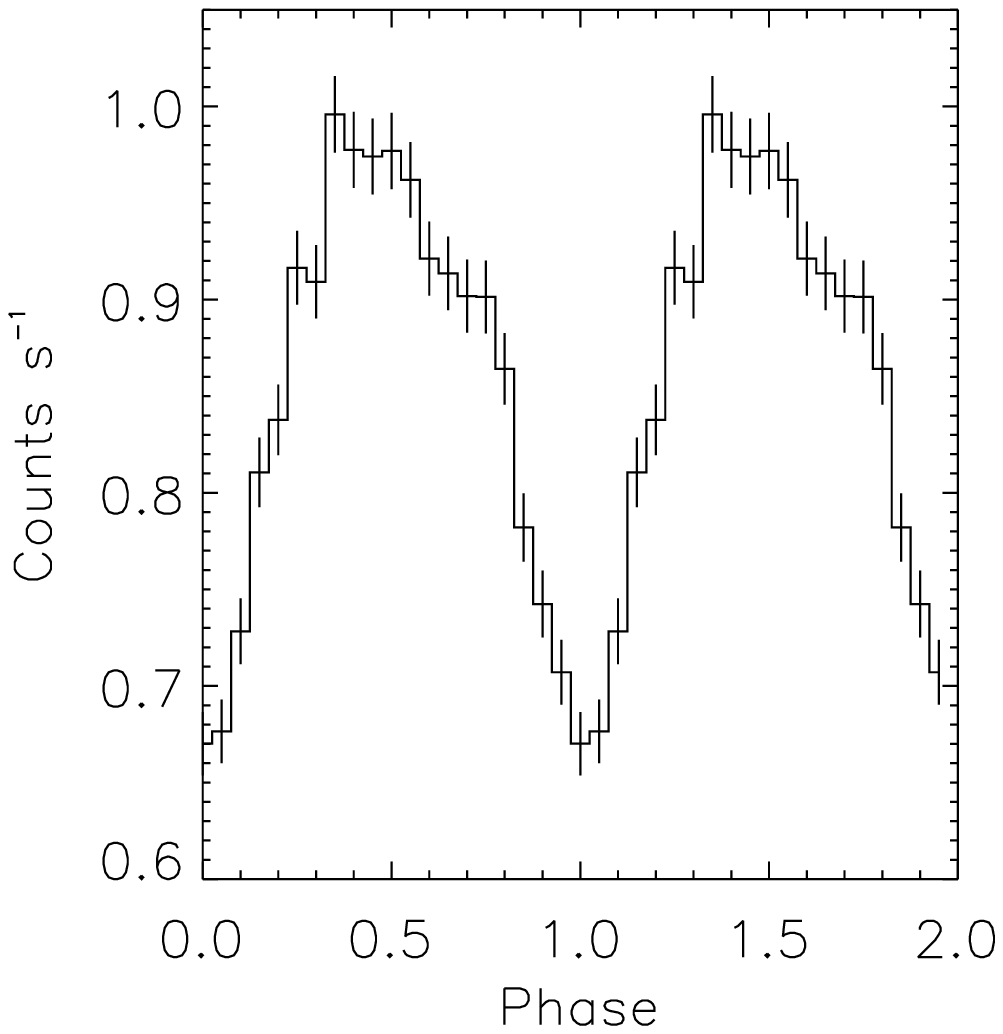,angle=0,width=5cm}
 }
\hbox{
 \psfig{figure=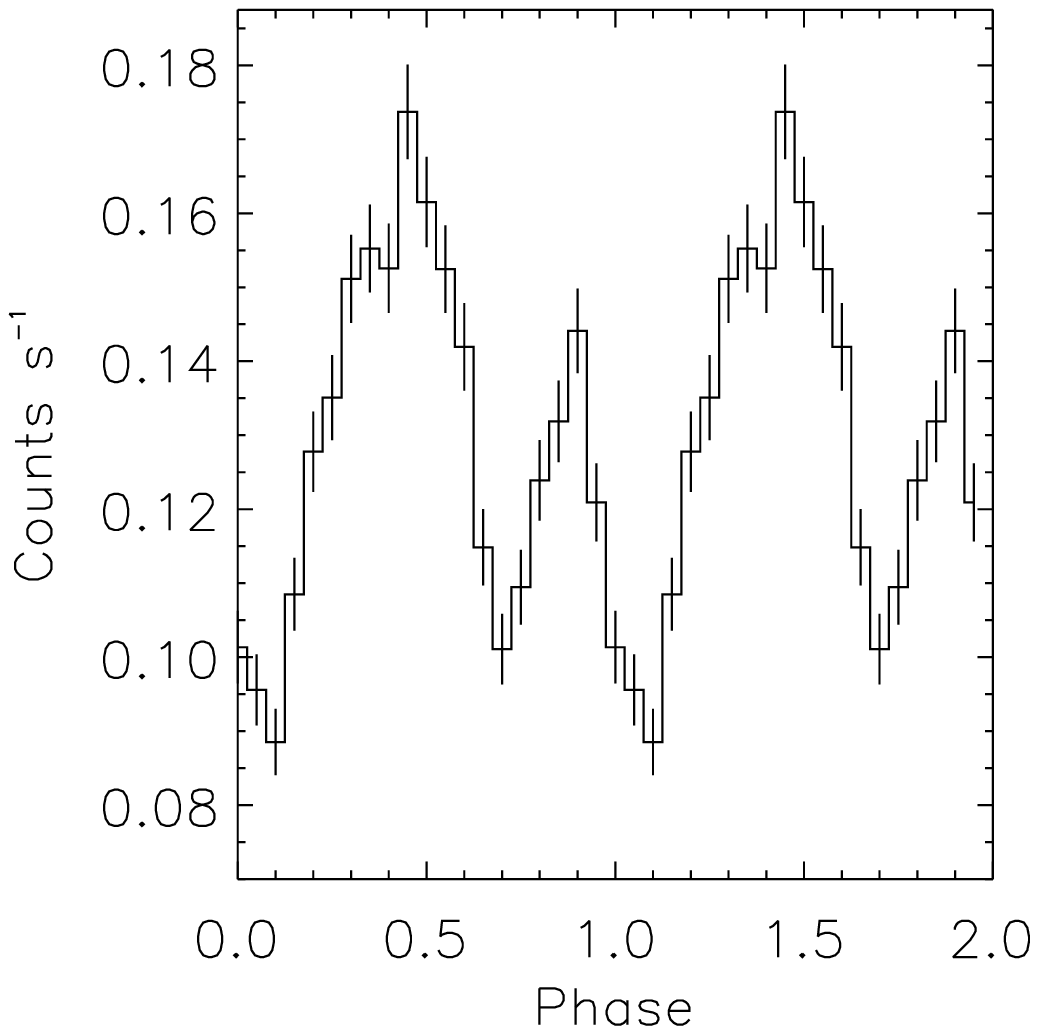,angle=0,width=5cm}
 \psfig{figure=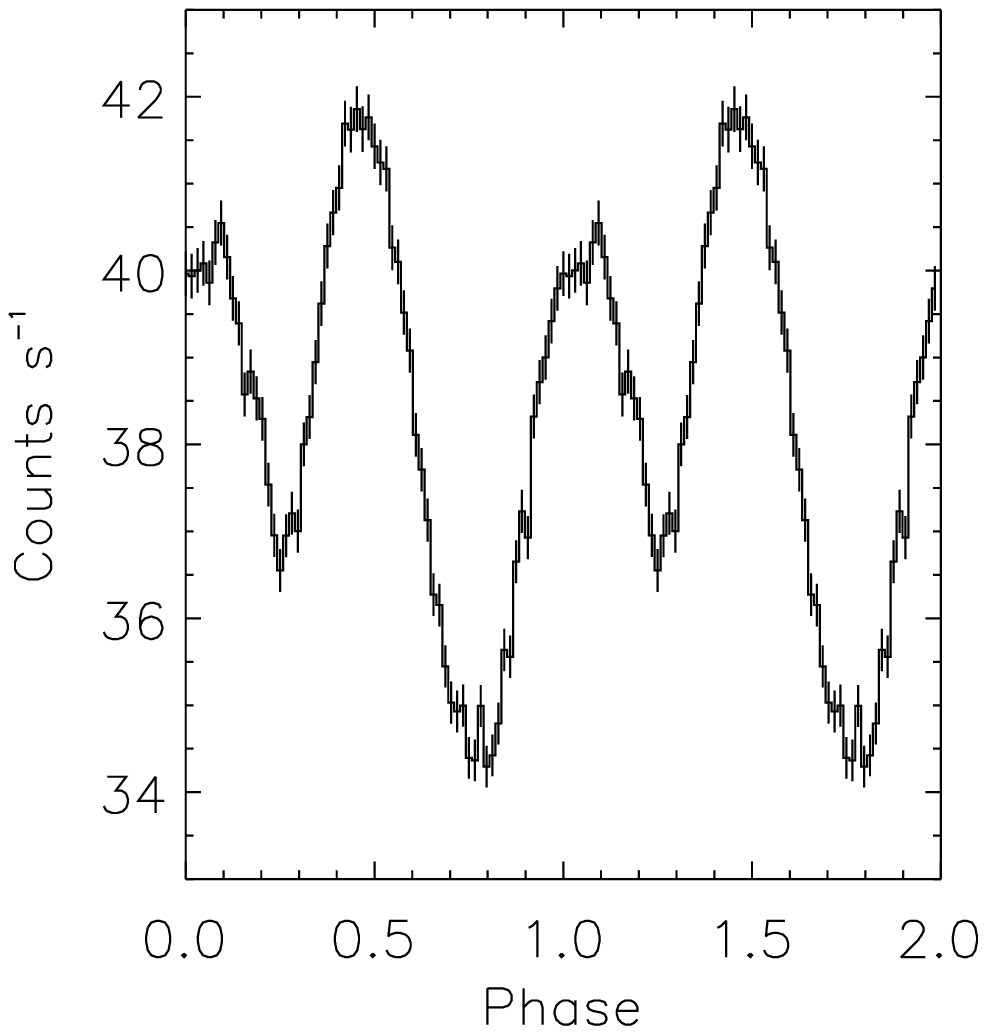,angle=0,width=5cm}
 }
}
 \caption{Pulse profiles of AXPs and SGRs obtained with the XMM-Newton EPIC instrument
 in the 1-10 keV band (courtesy P.Esposito). From top left to bottom right the sources are:
 \oo , \xte , \zerosei , \zerozero , \smc\ and \uu .}
\label{fig-LC}
\end{figure*}

Some representative pulse profiles in the soft X--ray range  are
plotted in Fig.~\ref{fig-LC}. Most magnetars have pulse profiles
consisting of a single broad peak of nearly sinusoidal shape,
while a few sources have double peaked profiles (e.g., \ee , \uu ,
\smc ). A large variety of pulsed fractions is also observed.  In
most objects the pulse profiles are energy dependent and also
change as a function of time. The time variations are more
dramatic in the case of the SGRs \citep{gog02}, which tend to have
more structured profiles when they are in periods of bursting
activity. Strong variations in the pulse profiles have been seen
to occur also on short timescales during the pulsating tails that
follow the giant flares, most likely due to large scale
rearrangements of the magnetic fields in the emitting regions.

In a few AXPs the pulsed fraction is smaller when  the X-ray flux
is higher. The best example  is \oo\ \citep{tie05a,tam07}, whose
pulsed fraction in the low state was $\sim$91\% (the highest of
any magnetar) and decreased to $\sim$20\% when the flux increased.
An anticorrelation between flux and pulsed fraction was also seen
during the outbursts of June 2002 in \ee\ \citep{woo04} and of
September 2006 in \cxo\ \citep{mun07}. The opposite behavior, i.e.
a decreasing pulsed fraction,  is instead seen during the long
flux decay of \xte\ \citep{got07b}. Finally, long term changes in
the pulsed fraction of \uu\ were  occurring while the overall flux
remained constant \citep{gon07}.

\subsection{Glitches}
\label{glit}

Glitches have been observed in practically all the AXPs for which
adequate timing data have been taken over sufficiently long time
periods (see Table \ref{tab-timing}). Most of the AXP glitch
properties are consistent with those of young radio pulsars
($\tau_c\sim10^3$-10$^5$ yr), thus giving independent evidence
that AXPs and SGRs are relatively young objects. However, their
glitch amplitude and frequency is larger than that seen in radio
pulsars of comparable spin periods, which exhibit smaller and more
rare glitches. This seems to suggest that the age of a neutron
star, rather than its rotation rate, is determining the glitch
properties.

The glitches in \rxs\ have different properties in their recovery
times \citep{dal03}, which are difficult to reconcile with a
single mechanisms, such as, e.g., the standard vortex unpinning
model. In particular, the recovery time after the largest glitch,
was considerably shorter than typically observed in radio pulsars,
and similar to that seen after the \ee\ glitch.

It seems that the variety of glitch properties in AXPs/SGRs can be
better explained in terms of  starquakes models. In magnetars,
localized starquakes are expected due to the stresses induced by
the magnetic field on the neutron star crust. The resulting
movements of the magnetic foot-points are also thought to generate
Alfv\'{e}n waves in the magnetosphere responsible for the short
burst.  The June 2002 event in \ee\ \citep{kas03}, when both
bursts and a glitch were observed, supports this scenario, while
the apparent lack of bursts associated with the glitches in \rxs\
might be due to the sparse coverage of the observations.

A large increase in the spin period was observed in connection
with the August 1998 giant flare of \zerozero , however the lack
of adequate timing measurements in the $\sim$2 months preceding
this event, does not allow to distinguish among different
interpretations \citep{woo99b}. It is possible that an
``anti-glitch'' (i.e. a step-like frequency \textit{decrease})
with $\Delta P/P$=10$^{-4}$, coincident in time with the giant
flare occurred due to a sudden unpinning of the neutron superfluid
vortex lines. This requires that, contrary to ordinary neutron
stars, the neutron superfluid in magnetars rotates more slowly
than the crust \citep{tho00}. A second possibility  \citep{pal02}
is that the giant flare was followed by a period lasting minutes
or hours with a spin-down larger by about two orders of magnitude
than the long term average value of $\sim8\times10^{-11}$ s
s$^{-1}$. Finally, it cannot be excluded that the source underwent
an increased spin-down in the two months preceding the flare. In
this respect it is interesting to note that no (anti-)glitches
were seen in the much more energetic giant flare of \zerosei\ and
that the same source exhibited significant $\pdot$ variations in
the months preceding the giant flare \citep{woo07}.

\begin{landscape}
\begin{table}[t]
\caption{Timing parameters AXPs and SGRs } \centering
\label{tab-timing}       
\begin{tabular}{lcccl}
\hline\noalign{\smallskip}

Name &   P  & $\pdot$           &  Glitches      & Bursts\\
     &  (s) &  (s s$^{-1}$)      &  $\Delta\nu/\nu$ &  \\

\tableheadseprule\noalign{\smallskip}

\smc         &  8.02    & 1.9 10$^{-11}$  &         & \\
             & \cite{lam03}  &  \cite{mcg05}  &                     &     \\
 \hline
\uu          &  8.69    & 2 10$^{-12}$  &       possibly one in 1998-2000? & six in 2006-2007\\
            & \cite{isr94}  & \cite{hel94,gav02b}       & \cite{dib07b} &   \cite{gav08}  \\
\hline
\oo         &  6.45       & (1--10) 10$^{-11}$        & March 2007 \cite{dib07c}  & two in 2001 \cite{gav02} \\
           &  \cite{sew86}   & \cite{cor90,mer95b,gav04b}    & $(2.7\pm0.7)\times10^{-6}$ & 29 June 2004 \cite{gav06} with long tail  \\
\hline
\qui      &  2.07     & 2.32 10$^{-11}$  &    &   \\
         & \cite{cam07c}   &  \cite{cam07c}          &  &   \\
\hline
\cxo    & 10.6    &  9.2 10$^{-13}$  &        21 Sept 2006 \cite{isr07b}  & 21 Sept 2006  \\
        & \cite{mun06} &   \cite{isr07b}       &   6$\times10^{-5}$     & \cite{isr07b}   \\
\hline\smallskip
\rxs     & 11.00    & 2.4 10$^{-11}$ &       several \cite{kas00,dal03,dib07a,isr07} & \\
        & \cite{sug97} & \cite{isr99}       &     6$\times10^{-7}$ -- 3$\times 10^{-6}$      &     \\
\hline
\xte     &  5.54     & (0.8--2.2) 10$^{-11}$      &   & four from 9/2003 to 4/2004  \\
         & \cite{ibr04}  &  \cite{ibr04,got05}    &      &  \cite{woo05}   \\
\hline
\kes     & 11.77   &  4.1 10$^{-11}$   &   three    \cite{dib07a}     & \\
        & \cite{vas97} &  \cite{got99} & 1.4$\times10^{-6}$ -- 5.6$\times 10^{-6}$  &     \\
\hline
\axj     &  6.97  &     &     &   \\
         &\cite{tor98} &      &          &     \\
\hline
\ee      &  6.98   &  4.8 10$^{-13}$  &   June 2002 \cite{woo04}    & $>$80 in June 2002 \\
        &  \cite{fah81} &  \cite{gav02b}   &    $(4.24\pm0.11)\times10^{-6}$  & \cite{kas03}    \\
\hline
 \hline
\lmc     &  8        & 6.5 10$^{-11}$ &        &  active in 1979-1983\\
         &  \cite{cli80}  &  \cite{kul03}    &         &     \\
\hline
\sedici  &        &   &            &    active in June-July 1998  \\
         &       &    &        &     \\
\hline
\zerosei &  7.6   &  (0.8--8) 10$^{-12}$      &            & several active periods \\
         & \cite{kou98} & \cite{kou98,woo07}             &         &  \\
\hline
\zerozero&  5.2       &  (5-14) 10$^{-11}$       & during giant flare \cite{woo99b}  & several active periods \\
         &  \cite{hur99e}  & \cite{kou99,woo99b} &  --10$^{-4}$  &     \\
\hline

\noalign{\smallskip}\hline
\end{tabular}
\end{table}
\end{landscape}


\subsection{Quasi Periodic Oscillations}
\label{qpo}

\begin{figure*}
\psfig{figure=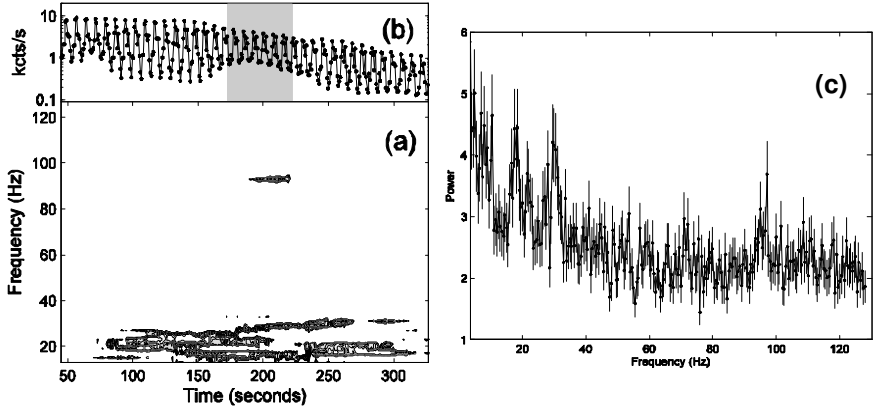,angle=0,width=12.5cm}
\caption{QPOs in the giant flare from \zerosei\ (from
\citet{isr05b}). The image in panel (a) shows a dynamical power
spectrum, where the frequencies of the detected QPOs can be seen
as a function of time. Panel (b) shows the light curve in the same
time interval of panel (a). Panel (c) shows the power spectrum
corresponding to the time interval 200--300 s; the peaks
corresponding to QPOs at $\sim$18, $\sim$30 and $\sim$95 Hz are
visible. } \label{fig-sgr1806_QPO}
\end{figure*}

A recent interesting result is the discovery of quasi-periodic
oscillations (QPO) in the decaying  tails of SGRs giant flares.
This phenomenon was discovered in RXTE data of the December 2004
very energetic giant flare from \zerosei\ \citep{isr05b}. QPOs at
a frequency of 92.5 Hz were present in a 50 s long interval,
corresponding to a bump in the unpulsed component of the X-ray
emission, about 200 s after the start of the flare (see
Fig.~\ref{fig-sgr1806_QPO}. They occurred only at a particular
phase of the 7.6 s neutron star spin period, away from the main
peak. Oscillations with a smaller significance, but lasting for a
longer time interval, were also detected at lower frequencies (18
and 30 Hz).

An independent confirmation of the 92.5 Hz and 18 Hz oscillations
in \zerosei\ was obtained with data from the RHESSI satellite,
which also showed  other QPOs at 26 Hz and, most remarkably, at
626.5 Hz in a different rotational phase and at higher energy
\citep{wat06}. Further analysis of the RXTE data of the same giant
flare \citep{str06} showed other time and pulse phase dependent
QPOs at $\sim$150, 625, 1480 Hz (lower significance QPOs were also
present at 720 and 2384 Hz).

The discovery of QPOs in \zerosei\ prompted a search for the same
phenomenon in the RXTE data of the August 1998  giant flare of
\zerozero . This led to the detection of QPOs at frequencies of
28, 54, 84 and 155 Hz \citep{str05}. The signal with the highest
rms amplitude (84 Hz) was visible only for one second. The other
QPOs lasted much longer ($\sim$90 s), and, similar to the 92.5 Hz
QPO of \zerosei , they were present only in a rotational phase
interval, the same of the 84 Hz oscillations. In retrospect, it is
also likely that the hint for a 43 Hz periodicity seen in the
March 1979 flare from \lmc\ \citep{bar83} was due to the same
phenomenon.

The QPOs observed in the tails of giant flares are most likely due
to seismic oscillations induced by the large crustal fractures
occurring in these extremely energetic events, similar to what
happens after earthquakes. The oscillations could be limited to
the crust or involve the whole neutron star, depending on the
unknown amount of core-crust coupling. The correct identification
of the observed vibrational modes is not obvious, with the excited
harmonics probably depending on the site and nature of the crustal
fracture. The theoretical models suggest that toroidal modes
should be the ones most easily excited \citep{dun98} and the
resulting horizontal displacements could easily couple with the
external magnetic field, causing the observed modulations in the
X-ray flux. Given that the mode frequencies depend on the neutron
star mass, radius, magnetic field,  composition and structure, the
QPO studies offer very interesting diagnostics, as is the case for
astroseismology. In principle, it might even be possible to obtain
some constraints on the neutron star equation of state.

Unfortunately, such studies are made difficult by the rarity and
by the unpredictable occurrence of SGR giant flares, as well as by
the fact that we do not observe directly the crust vibrations, but
only their effect on the X-ray emission, mediated by the magnetic
field. This is testified by the sporadic nature of the observed
signals and their connection to different rotational phases,
probably reflecting the complex geometry of the magnetic fields
and radiation beam patterns.

\section{Counterparts at long wavelengths}

\subsection{Optical and Infrared}
\label{optical}

Much progress has been done in the search for optical/IR
counterparts. Currently, counterparts have been securely
identified for five magnetars and promising candidates have been
detected for  most of the remaining ones, thanks to the detection
of objects showing variability or unusual colors inside the small
error regions obtained with Chandra (and in some cases from radio
observations,  see Table~\ref{tab-positions}).

\begin{figure*}
\hbox{ \psfig{figure=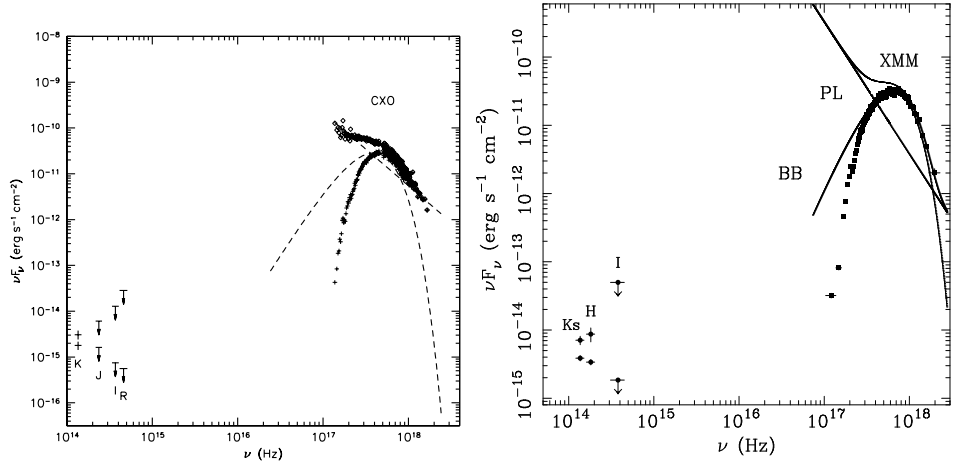,angle=0,width=6cm}
\psfig{figure=xtej1810_X_IR_isr04.eps,angle=0,width=7cm} }
\caption{Broad band spectra of \ee\ (left panel,  from
\citet{hul01}) and \xte\ (right panel, from \citet{isr04a}). The
different values for the optical/IR data refer to absorbed and
unabsorbed values. }
 \label{fig-X-IR}
\end{figure*}

All the (candidate) counterparts are very faint (see Table
\ref{tab-ctpt}), giving ratios of the X-ray to IR flux larger than
a few thousands. This excludes in most cases the presence of
normal stars. The IR fluxes lie well below the extrapolation of
the steep power laws often used to fit the soft X-ray spectra, but
above the the extrapolation of the X-ray blackbody components
(Fig.~\ref{fig-X-IR}).

After the June 2002 outburst the IR counterpart of \ee\ was a
factor $\sim$3-4 brighter than the ``quiescent'' level
\citep{kas03}. The IR and X-ray fluxes subsequently decayed in a
similar way, suggesting a close link between the emission
processes in these two energy ranges. This was interpreted as
evidence for a non-thermal, magnetospheric origin of the IR
radiation \citep{tam04}, but \citet{ert06a} showed that the data
can also be explained as emission from a residual disk pushed away
by an energetic flare and gradually relaxing back to its original
configuration.

Long term IR  variability has been reported also for \xte\
\citep{rea04}, \rxs\ \citep{dur06} (but see \citep{tes07}), and
\zerosei\ \citep{isr05}. \citet{hul04} found long term variability
in the IR flux of \uu, but not in the optical\footnote{However,
further observations showed that also the optical flux varies
\citep{dur06}.}.  Thus it seems that, similarly to what happened
in the X-ray range, variability is detected whenever repeated
accurate measurements are available. Correlations between IR and
X-ray flux variations have been searched for, but a single
coherent picture has not been found yet. A positive correlation
was reported for flares and transient outbursts (e.g. \ee,
\citep{kas03}; \xte, \citep{rea04}). In other cases the situation
is more complex, and the sparse coverage of the observations does
not allow to derive firm conclusions. For example, IR variations
on  a timescale of days have been  reported for \uu\
\citep{dur06c}, but no simultaneous X-ray data exist, and \xte\
showed  fluctuations in the IR flux uncorrelated with the X-ray
decay \citep{tes07,cam07e}.

These results point to possibly different origins for the X--ray
and (optical)/IR emission. For example, the latter could be non
thermal coherent emission from plasma instabilities above the
plasma frequency \citep{eic02}, in which case it would be probably
pulsed and polarized. Unfortunately most counterparts are too
faint in the NIR to test these predictions.

\uu\ is the only magnetar securely detected in the optical band,
and the only one showing optical pulsations \citep{ker02,dhi05}.
The optical pulses have the same period and approximate phase of
the X-rays, but a larger pulsed fraction (see also
sect.~\ref{other}). Possible models to explain the pulsed optical
emission from \uu\ are discussed in \cite{ert04}.


\begin{table}[t]
\caption{Optical and Infrared Counterparts or upper limits}
\centering
\label{tab-ctpt}       
\begin{tabular}{lll}

\hline\noalign{\smallskip}

Name &   Counterparts & Comments  and references\\[3pt]

\tableheadseprule\noalign{\smallskip}

\smc      &                         & V\gtsima26  \\
          &                         &  \citep{dur08} \\
          & & \\
\uu       & K = 19.7--20.8            & variable,  optical pulsed    \\
          & R=24.9--25.6            & \cite{hul00,ker02,dhi05}\\
          & & \\
\oo       & K= 19.4--21.5              & variable   \\
          & \textit{r}$'>$25.6              &  \citep{wan02,wan07b} \\
          & & \\
\qui     &     &   K\gtsima17.5  \\
     &    & \citep{gel07}\\
        & & \\
\cxo      &                         &  K \gtsima 21\\
          &                         & \cite{wan06b}            \\
          & & \\
\rxs      &                          & several candidates K= 18.9--19.3 \\
          &                          & \citep{dur06,tes07} \\
          & & \\
\kes      &                           & several candidates (K=18--21)  \\
          &                           & one variable \citep{mer01,dur05,tes07} \\
          & & \\
\xte      & K= 20.8--21.4              &  variable  \\
          &                           &  \citep{isr04a,rea04,tes07,cam07e} \\
          & & \\
\axj      &                           & H$>$21  \\
          &                           & \citep{isr04b} \\
          & & \\
\ee       & K$_s$ = 21.7--20.4          & variable (brighter after June 2002 outburst) \\
          & R$>$26.4                     & \cite{hul01,tam04}\\
          & & \\
\lmc        & & \citep{kap01} \\
          &  & \\
          & & \\
\sedici   &                            &  K \gtsima 20 \citep{wac04} \\
          &   & \\
          & & \\
\zerosei  & K=19.3--22              & variable \\
          &                                & \citep{isr05,kos05} \\
          & & \\
\zerozero  &        & variable candidate \\
           &        & K$\sim$19.7 \cite{tes07,kap02a}\\
          & & \\
\noalign{\smallskip}\hline
\end{tabular}
\end{table}


\subsection{Pulsed radio emission}
\label{radio}

Early observations to search  for (pulsed) radio emission from
AXPs and SGRs gave negative results\footnote{Transient radio
emission has been observed after the two giant flares of
\zerozero\ \citep{fra99} and \zerosei\ \citep{gae05,cam05}. This
emission is thought to originate from shocks in mildly
relativistic matter ejected during the giant flares
\citep{gra06}.}, although the luminosity limits were above those
of many weak radio pulsars \citep{gae01}. It was initially
believed that the absence of radio emission was a distinctive
characteristic of magnetars. This is a natural expectation in
models based on accretion, that would quench any radio pulsar
mechanism. For the magnetar model, it was suggested that photon
splitting in the high magnetic field could dominate over pair
creation, thus suppressing the charged particle cascades that are
at the origin of the radio emission \citep{bar98,bar01}. However,
photon splitting applies only to one polarization mode: photons of
the other mode cannot split. Therefore this argument does not
apply, as also demonstrated by the existence of radio pulsars with
inferred dipole fields of several 10$^{13}$ G \citep{cam00,mcl04}.

A  point like radio source associated to the transient \xte\ was
discovered in 2004, about one year after the start of the X--ray
ouburst \citep{hal05b}, and later shown to consist of bright ($>$1
Jy), highly linearly polarized pulses at the neutron star rotation
period \citep{cam06}. The source was undetected in previous radio
data, obtained before the onset of the X--ray outburst. Recently,
radio pulsations at 2.07 s have been reported from \qui\
\citep{cam07c}, thus confirming the magnetar nature of this X--ray
source through a measurement of its spin-down. The fact that also
this object is a transient \citep{gel07} suggests that the
mechanisms responsible for the pulsed radio emission in magnetars
might be related to their transient nature. However, no radio
pulsations were seen in the other transient AXP, \cxo , after its
September 2006 outburst \citep{bur06b}, nor in \oo\ after the flux
enhancement accompanied by a glitch that occurred in March 2007
\citep{cam07d}. Deep searches for radio pulsations in persistent
AXPs have so far given negative results \citep{bur06}.

The radio properties of the two AXPs showing radio pulsations
differ in several respects from those of radio pulsars: their flux
is highly variable on daily timescales, their spectrum is very
flat with $\alpha>$--0.5  (where $S_{\nu} \propto \nu^{\alpha}$),
and their average pulse profile changes with time
\citep{cam07a,cam07c,cam08}. Such differences probably indicate
that the radio emitting regions are more complex than the dipolar
open field lines along which the radio emission in normal pulsars
is thought to originate.

\section{The magnetar model}
\label{magnetar}

\subsection{Formation and evolution of magnetars}
\label{formation}

The effects of a turbulent dynamo amplification occurring either
in a newly born, differentially rotating proto neutron star, or in
the  convective regions of its progenitor star, have been studied
in detail by \citet{tho93}. They concluded that very high magnetic
fields, in principle up to 3$\times$10$^{17}\times$(1
ms/P$_{\circ}$) G, can  be formed through an efficient dynamo  if
the neutron stars are born with sufficiently small periods, of the
order of P$_{o}$$\sim$1-2 ms, and if convection is present.
Population studies of radio pulsars indicate that such fast
initial periods are not common, and the birth spin periods
inferred from a few young pulsars are  of the order of a few tens
of milliseconds \citep{fau06}. However, plausible mechanisms have
been put forward that could lead to very high rotational speeds at
least for a small fraction of the neutron star population. Rapid
neutrino cooling in the proto neutron star is essential in driving
the strong turbulent convection which amplifies the seed field.
Such a dynamo operates only for $\sim$10 seconds, but is able to
generate fields as strong as 10$^{16}$ G, most likely with a
multipolar structure.

The dynamo responsible for the high magnetic field generation
requires that magnetars be born with very short rotation periods.
This formation scenario was predicted to have the two
observational consequences discussed below: \textit{(a)} magnetars
could have large spatial velocities, of the order of $\sim10^{3}$
km s$^{-1}$ and \textit{(b)} their associated supernovae should be
more energetic than ordinary core collapse supernovae
\citep{dun92}.

\textit{(a)} The combination of high magnetic field and very rapid
rotation is expected to impart a high velocity to the neutron
star, owing to the occurrence of several possible effects, like
anisotropic neutrino emission, magnetic winds, and mass ejection
due, e.g., to gravitational radiation instabilities. However,  up
to now, the observational evidence for large spatial velocities in
SGRs and AXPs is poor. The only measured proper motion has been
obtained with radio VLBA observations of \xte\ \citep{hel07}, and
corresponds to a transverse velocity of $\sim$180 (d/3 kpc) km
s$^{-1}$.  Most of the previously suggested associations with SNRs
are now considered chance coincidences (sec. \ref{snr}), the
exceptions being the three cases where the AXP is at the remnant
center and thus no high proper motion is required. The
identification of possible birthplaces in massive star clusters
(see section \ref{ass}) requires spatial velocities at most of a
few hundreds km s$^{-1}$, similar to those of radio pulsars.

\textit{(b)} A large fraction of the rotational energy of a newly
born magnetar, a few 10$^{52}$ erg, is lost due to the strong
magnetic braking. The initial spin-down  occurs on a timescale
$\sim$0.6 B$_{15}^{-2}$ (P$_o$/1 ms)$^{2}$ hours, shorter than the
supernova breakout time. Therefore, this additional  injected
energy should be reflected in the properties of the supernova
remnant. However, an estimate of the explosion energy of the
remnants containing magnetars \citep{vin06} yields values close to
the canonical supernova explosion energy of 10$^{51}$ erg,
implying initial periods longer than 5 ms.

The fact that these two predictions, high neutron star velocities
and energetic remnants, do not seem to be fulfilled, although
clearly  not sufficient to dismiss the dynamo formation mechanism,
has led some support to other formation scenarios. One alternative
theory is based on magnetic flux conservation arguments and
postulates that the distribution of field strengths in neutron
stars (and white dwarfs) simply reflects that of their
progenitors. In this ``fossil field'' scenario, the magnetars
would simply be the descendent of the massive stars with the
highest magnetic fields. The wide distribution of field strengths
in magnetic white dwarfs is thought to result from the spread in
the magnetic fields of their progenitors. Extrapolating this
result to the more massive progenitors of neutron stars could
explain the origin of magnetars \citep{fer06}. On average, higher
magnetic fluxes are expected in the more massive progenitors. The
evidence for a massive progenitor for the AXP \cxo\ in the open
cluster Westerlund 1 \citep{mun06}, and the young clusters of
massive stars found close to the locations of the SGRs
(sect.~\ref{ass}), seem to support this scenario.

Another possible evidence that high magnetic fields might also be
present in neutron stars born with relatively long spin periods
comes from the unusual X-ray source in the supernova remnant RCW
103, if its suggested interpretation in terms of a strongly braked
magnetar is confirmed (see sect.~\ref{rcw}).

\subsection{Origin of persistent emission and bursts}
\label{emission}

Young magnetars undergo a rapid spin down due to their strong
magnetic dipole radiation losses, reaching periods of several
seconds in a few thousands years. This rapid evolution toward the
so called ``death-line" in the B-P diagram explains why no
magnetars are observed at short rotational periods, and possibly
why they are not active in radio like normal  pulsars (but see
sect. \ref{radio}). Shortly in their life, magnetars slow down to
the point that their magnetic energy, E$_{mag}$$\sim10^{47}
(B/10^{15} G)^{2} (R/10$ km$)^{3}$ $\sim$ 10$^{46}$ (P/5 s)
($\pdot$/10$^{-11}$ s s$^{-1}$) ergs, is much larger than their
rotational energy. Such a huge energy reservoir is sufficient to
power for $\sim$10$^{4}$ years the persistent X--ray emission. The
giant flares sporadically emitted by SGRs, during which up to
$\sim10^{46}$ ergs can be released, are energetically more
challenging. This obviously limits the number of such events that
a magnetar can emit in its lifetime.

Different possibilities have been proposed to explain the observed
X--ray emission at a level of several $\sim10^{35}$ erg s$^{-1}$.
Magnetic field decay can provide a significant source of internal
heating. While ohmic dissipation and  Hall drift dominate the
field decay respectively in weakly (\ltsima10$^{11}$ G) and
moderately magnetized ($\sim10^{12-13}$ G) neutron stars, the most
relevant process in magnetars is ambipolar diffusion, which has a
characteristic  timescale t$_{amb}\sim10^4$ $\times$
($\frac{B_{core}}{10^{15} G}$)$^{-2}$ years \citep{tho96}. This
internal heating source yields a surface temperature higher than
that of a cooling neutron star of the same age and smaller
magnetic field. Furthermore, the enhanced thermal conductivity in
the strongly magnetized envelope, contributes to increase the
surface temperature \citep{hey97a,hey98a}.

The motion of the magnetic field, as it diffuses out of the
neutron star core, can also generate multiple small scale
fractures in the crust \citep{tho96}. This persistent seismic
activity produces low amplitude Alfv\'{e}n waves in the
magnetosphere, which can contribute to the X-ray emission, e.g.,
through particle acceleration leading to Comptonization and
particle bombardment of the surface. Stronger and less frequent
crust fractures provide a possible explanation for the short
bursts.

Persistent emission in magnetars can also be induced by the
twisting of the external magnetic field caused by the motions of
the star interior, where the magnetic field is dominated by a
toroidal component larger than the external dipole. The twisting
motion of the crust sustains steady electric currents in the
magnetosphere, which provide an additional source of heating for
the star surface \citep{tho00}.

In the context of the magnetar model two mechanisms have been
proposed to describe the properties of transient magnetars: deep
crustal heating \citep{lyu02} and currents in the twisted
magnetosphere \citep{tho02}. The first model considers the effects
that a relatively fast energy deposition in the neutron star
crust, due for example to a sudden fracture or a gradual plastic
deformation, has on the surface thermal emission. This model was
studied primarily in connection with the ``afterglows'' observed
after giant and intermediate flares of SGRs, but was also applied
to flux decays seen on longer timescales, such as in \sedici\
\citep{kou03}. The time dependence of the surface ``thermal echo''
depends primarily on the thermal properties of the outer crust, as
well as on the depth of the energy deposition. Detailed modelling
of the observed light curves might thus lead to important
information on the star structure. However, the available
observations are far from showing a uniform picture and often
subject to  uncertainties that do not allow an easy comparison
with the models predictions, as well exemplified  \citep{mer06a}
by the case of \sedici\ shown in Fig.~\ref{fig-sgr1627_lcurve}.

The radiative mechanisms responsible for the bursts and flares,
which  in the magnetar model are explained in terms of magnetic
reconnections \citet{lyu02}, are extensively discussed in
\citet{tho95}. The short, soft bursts can be triggered by cracking
of the crust caused by the strong magnetic field. The crust
fractures perturb the magnetosphere and inject fireballs. The
bursts duration is dictated by the cooling time, but it depends
also on the vertical expansion of surface layers \citep{tho02}
and/or depth of heating \citep{lyu02}.

A different explanation for the  bursts origin has been proposed
in the ``fast-mode breakdown'' model \citep{hey05a},  in terms of
quantum electrodynamics processes occurring in magnetic fields
larger than B$_{QED}$. Also in this model, Alfv\'{e}n waves
induced by the crust motion are injected in the magnetosphere and
develop discontinuities similar to hydrodynamic shocks due to the
vacuum polarization. The wave energy is dissipated through
electron-positron pair production and the formation of optically
thick fireballs in the magnetosphere, which radiate mostly thermal
emission in the hard X-ray / soft gamma-ray range.

\subsection{Evidence for high magnetic fields}
\label{Bevidence}

The secular spin-down measured in magnetars allows to infer their
magnetic field through the  dipole braking relation B =
3.2$\times$10$^{19} (P \pdot)^{1/2}$ G. This yields values in the
range $\sim$(0.5--20)$\times10^{14}$ G. However, these estimates
are subject to some uncertainties since other plausible processes,
such as for example the ejection of a relativistic particles wind
\citep{har99}, can contribute to the torques acting on these
neutron stars. Up to now, attempts to estimate the magnetic field
strength through the measurement of cyclotron resonance features,
as successfully done for accreting pulsars, have been inconclusive
(sect.~\ref{lines}).

The most compelling evidence for the presence of high magnetic
fields comes from the extreme properties of the giant flares
observed in SGRs (sect.~\ref{gf}). The first object to be
interpreted as a magnetar was in fact \lmc, responsible for the
exceptional giant flare observed on 1979 March 5  \citep{maz79}.
Several properties of this event could naturally be explained by
invoking a super strong magnetic field \citep{dun92,pac92}. The
extremely challenging properties of this first observed giant
flare were subsequently confirmed by the more detailed
observations of similar events from two other SGRs.

Two aspects of the March 1979 event were crucial for the magnetar
interpretation: its spatial coincidence with the young supernova
remnant N49 in the Large Magellanic Cloud, which immediately
enabled to set the energetics through a secure distance
determination, and the evidence for a periodicity of 8 s, strongly
hinting to the presence of a rotating neutron star. As discussed
above (sect. \ref{gf}), giant flares are characterized by an
initial hard spike of emission up to the MeV range, lasting a
fraction of a second, followed by a long tail (several minutes)
with a softer spectrum and clearly showing the periodic modulation
due to the neutron star rotation. Magnetic confinement of the hot
plasma responsible for the pulsating tails is one of several
evidences for the presence of a high field, and sets a lower limit
of the order of a few 10$^{14}$ G.

Other motivations for a high magnetic field include: (a) the
reduction, due to the magnetic field, in the photon opacity
required to exceed by at least a factor $\sim10^{3}$ the Eddington
limit for a neutron star in the soft $\gamma$-ray bursts; (b) the
necessity of providing enough magnetic free energy to power the
giant flares; (c) the short duration of the initial spikes,
consistent with the propagation with Alfv\'{e}n speed of the
magnetic instability over the whole neutron star surface
\citep{tho95}.

A strong dipole field also provides a natural way to slow-down a
neutron star to a long period within a relatively short time. In
the case of \lmc, currently spinning at 8 s,  the associated SNR
implies an age of $\sim10^{4}$ yrs. Although most of the  proposed
associations of the other magnetars with SNRs are no more
considered significant (sect. \ref{ass}), their small scale height
on the Galactic plane and their tendency to be found in regions of
active star formation and close to clusters of very massive stars
\citep{cor04,vrb00,klo04,mun06} indicate that magnetars are young
objects.

Finally,  an independent evidence for superstrong magnetic fields
in SGRs has been recently pointed out by \citet{vie07} who
considered the high frequency QPOs observed in the giant flare of
\zerosei\ (sect.~\ref{qpo}). The 625 and 1840 Hz QPOs involve
extremely large and rapid luminosity variations, with
$\Delta$L/$\Delta$t as large as several 10$^{43}$ erg s$^{-2}$
(the exact value depends on the assumed beaming). This value
exceeds the Cavallo-Rees luminosity-variability limit
$\Delta$L/$\Delta$t $<$ $\eta$~2$\times10^{42}$  erg s$^{-2}$,
where $\eta$ is the efficiency of matter to radiation conversion
\citep{cav78}.  The relativistic effects, generally invoked to
circumvent this limit (e.g. in blazars and gamma-ray bursts) are
unlikely to be at work in the SGR QPO phenomenon. \citet{vie07}
instead propose that the Cavallo-Rees limit does not apply due to
the reduction in the photon scattering cross section induced by
the strong magnetic field. In this way a lower limit of
$\sim2\times10^{15}$ G $(10~\textrm{km} / R_{NS})^3$
$(0.1/\eta)^{1/2}$ for the surface magnetic field is derived.

\subsection{Twisted magnetospheres}
\label{twist}

\begin{figure*}
\psfig{figure=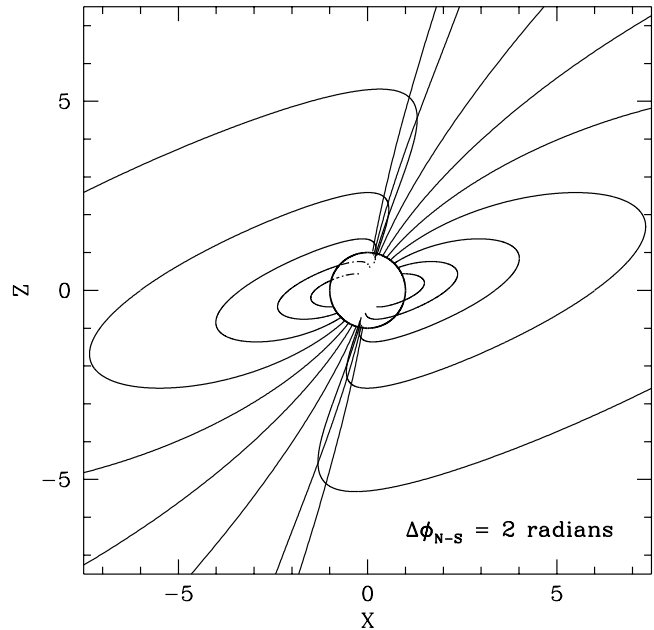,angle=0,width=9cm} \caption{Example
of a twisted dipole magnetic field (from \citet{tho02}). The twist
angle between the northern and southern hemisphere is $\Delta
\Phi_{N-S}$= 2 rad. Dashed lines indicate the part of the field
lines behind the neutron star. } \label{fig-twist}
\end{figure*}

\citet{tho02} studied the properties of twisted magnetospheres
threaded by large scale electrical currents. It is believed that
the magnetar internal field is tightly wound up in a toroidal
configuration and is up to a factor $\sim$10 stronger than the
external field. The unwinding of the internal field shears the
neutron star crust. The rotational motions of the crust  provide a
source of helicity for the external magnetosphere by twisting the
magnetic fields which are anchored to the star surface (see
Fig.~\ref{fig-twist}). A globally twisted magnetosphere, instead
than a simple dipolar configuration, could be the main difference
between magnetars and high B radio pulsars.

\begin{figure*}
\psfig{figure=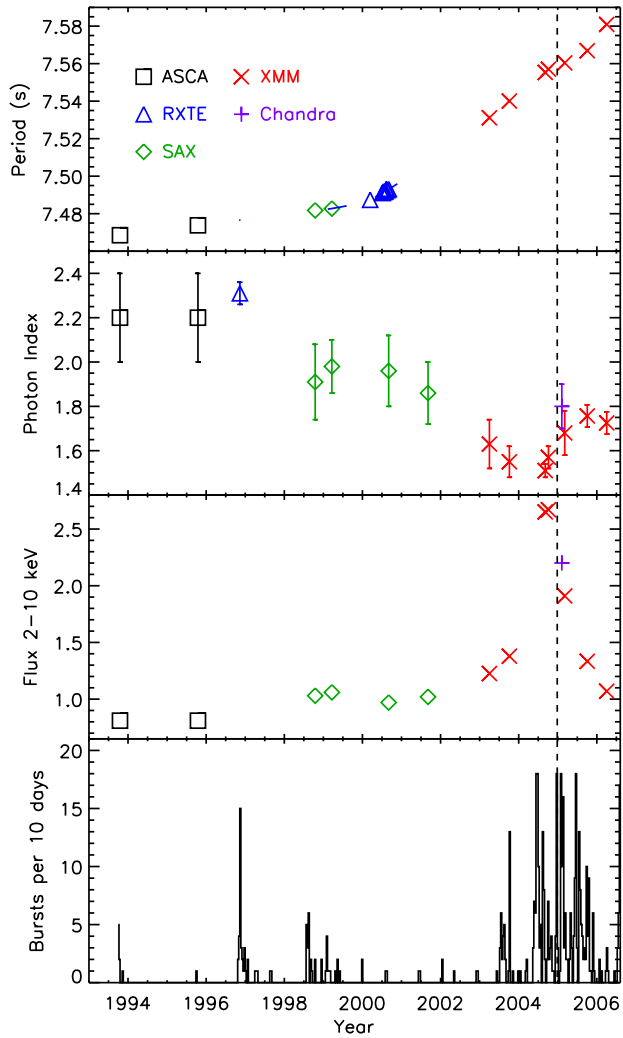,angle=0,width=9cm}
\caption{Evolution of the properties of \zerosei\ before and
immediately after the December 2004 giant flare. From top to
bottom: pulse period, power-law photon index, X--ray flux, rate of
bursts. The vertical line indicates the date of the giant flare.
In the years preceding the giant flare the spin-down increased,
the spectrum hardened, the X--ray flux and bursting activity
increased (see \citep{mer05c} for details).}
 \label{fig-sgr1806_history}
\end{figure*}

The presence of a twisted magnetosphere ($B_\phi\neq 0$) has
several interesting consequences. A twisted, force-free
magnetosphere supports   electrical currents several orders of
magnitude larger than the Goldreich-Julian current flowing along
open field lines in normal pulsars. The strong flow of charged
particles heats the neutron star crust and produces a significant
optical depth for resonant cyclotron scattering in the
magnetosphere. Repeated scattering of the thermal photons emitted
at the star surface can give rise to significant high-energy
tails. The optical depth is proportional to the twist angle, thus
a spectral hardening is expected when the twist increases. Another
consequence of the twisted field is that the spin-down torque is
larger than that of a dipolar field of the same strength. Given
that both the spectral hardening and the spin-down rate increase
with the twist angle, a correlation between these quantities is
expected. In fact the presence of such a correlation has been
reported by \citet{mar01}. Since the stresses building up in the
neutron star crust lead to crustal fractures which are at the base
of the burst emission, it is also expected that a twist angle
increase should give rise to an enhanced bursting activity. The
overall evolution of \zerosei\ in the years preceding the giant
flare of December 2004 (see Fig.~\ref{fig-sgr1806_history}) seems
to support these predictions \citep{mer05c}.

The magnetar starquakes and twisting magnetic field lead to the
formation of an electron/positron corona in the closed
magnetosphere. The corona consists of closed flux tubes,  anchored
on both ends to the neutron star surface and permeated by currents
driven by the twisting motion of their footpoints. The persistent
hard X--ray emission extending up to $\sim$100 keV originates in a
transition layer between the corona and the atmosphere, while the
optical and IR are emitted by curvature radiation in the corona
\citep{bel07}. The gradual dissipation of the magnetospheric
currents can also provide plausible mechanisms for the generation
of persistent soft $\gamma$-ray emission \citep{tho05}.

Recently, several studies concentrated on the derivation of
theoretical spectral models for magnetars. \citet{lyu06b} derived
a semi-analytical model to account, in a one-dimensional
approximation,  for the effects of multiple resonant scatter in
the magnetosphere on the blackbody emission from the magnetar
surface. Their model provides a good fit to a typical AXP spectrum
in the 1-10 keV range \citep{rea07c}.  A detailed 3-D Monte Carlo
simulation has been instead carried out by \citet{fer07}. Their
models are quite successful to reproduce spectra and pulse
profiles of AXPs in the 1-10 keV range with broad and mildly
relativistic particle distributions and twist angles of
$\sim$0.3-1 rad, while they over predict the thermal components of
SGRs.

\subsection{Hard X-ray tails}

In the context of the twisted magnetosphere model, two
possibilities have been proposed to explain the high-energy
emission from magnetars \citep{tho05}: \textit{(a)} bremsstrahlung
from a thin turbulent layer of the star's surface, heated to
kT$\sim$100 keV by magnetospheric currents, and \textit{(b)}
synchrotron emission from pairs produced at a height of $\sim100$
km above the neutron star. In the first case a cut-off at a few
hundred keV is expected, while in the second case the spectrum
should extend to higher energies, peaking around one MeV. The
currently available data are insufficient to discriminate between
the two cases by measuring the energy of the spectral cut-off,
which is required to avoid exceeding the upper limits obtained
with the Comptel instrument in the few MeV region.

Resonant cyclotron scattering  is thought to play an important
role in the production of  hard X--ray emission from magnetars
\citep{bar07}. In strong magnetic fields the Compton scattering is
resonant at the cyclotron energy, with a cross section much higher
than the Thomson value. The surface thermal photons (kT$\sim$keV)
propagating outward will at a given radius scatter resonantly,
i.e. they are absorbed and immediately re-emitted, if there is
plasma in the magnetosphere. While in normal pulsars the plasma
density is too small to produce a high optical depth, this is not
the case in magnetars, which have charges, with a density much
higher than the Goldreich-Julian one, flowing in their
magnetospheres. These charges could be accelerated along open
field lines (as in radio pulsars) or they could be due to the
large scale currents that are thought to be present in twisted
magnetospheres. In this case they also permeate the closed field
lines. Repeated scatterings of the surface thermal photons produce
the hard tails. Geometry effects are important here, as well as
the fact that the scattering region is large and hence the
magnetic field is not homogeneous. The transmitted flux is made by
the photons that on average gain energy. It is interesting to note
that the scattering plasma does not need to be highly
relativistic. The photon energy increase is taken at the expense
of the electrons. Thus, in the case of currents in twisted
magnetospheres the ultimate energy source is still the magnetic
field.

According to \citet{hey05b} the hard X-ray emission could instead
be due to synchrotron radiation. These authors showed that the
fast-mode breakdown model they developed to explain the bursts
(sect.~\ref{emission}), also predicts the presence of a
non-thermal distribution of electrons and positrons in the outer
parts of the magnetosphere. The quiescent hard X-ray emission, and
possibly also the optical/IR, would be associated to small scale
crust shifts generating fast modes whose breakdown is insufficient
to produce the fireballs responsible for the bursts.

\section{ALTERNATIVE MODELS}
\label{other}

\subsection{Accretion from fossil disks}
\label{disk}

Several proposals to explain the properties of the AXPs (and to a
lesser extent of SGRs) are based on  isolated neutron stars
surrounded by residual disks. In this class of models, that in
general do not require particularly high magnetic fields,  the
presence of a disk is invoked to account for the rapid spin-down,
but different mechanisms for its formation and different origins
for the observed X-ray luminosity have been considered.

It has been proposed that the AXP could be one possible outcome of
the common envelope evolutionary phase of close high mass X-ray
binaries, with residual accretion  disk forming after the complete
spiral-in of a neutron star in the envelope of its giant companion
\citep{van95,gho97}. For residual disks masses of about
0.01$\msun$, initial spin periods of 2-50 ms, and magnetic fields
in the upper range of the distribution of normal pulsars
($\sim10^{13}$ G), the propeller torques can spin down the neutron
star to  periods of a few seconds in less than 10$^4$ years
\citep{cha00a}. After the propeller phase, the neutron star can
start to accrete significantly, becoming visible as an AXP with a
period close to the equilibrium value, which would slowly increase
owing to the decreasing accretion rate in the disk. In this model
the upper cut off in the AXP period distribution is explained by
invoking a significant drop in the accretion efficiency due to an
advection dominated flow when the accretion rate falls below
$\sim$0.01 of the Eddington value \citep{cha00b}.

\citet{alp01} suggested that the properties of a fall-back disk
are among the fundamental parameters, together with initial spin
period and magnetic field, that determine the fate of newly born
neutron stars. He proposed a scenario which attempts to unify the
different classes of isolated neutron stars: radio pulsars, AXPs
and SGRs, Compact Central Objects in SNRs  and X-ray Dim neutron
stars (see sect.~\ref{ins} for a discussion of these objects).

According to \citet{mar01b} the formation of disks around SGRs and
AXPs is favored because, compared to normal radio pulsars, they
are born in denser interstellar medium regions and have larger
spatial velocities. Their supernova remnants expansion are
expected to rapidly decelerate through interaction with the dense
environment, thus forming strong reverse shocks that would push
back part of the ejecta toward the neutron star. In addition, high
velocity neutron stars might be nearly co-moving with the
supernova ejecta, favoring their capture. In this model the
different observational properties of AXPs and SGRs are ascribed
to their birth location rather than to an intrinsic difference
with the other neutron stars. However, the evidence for different
birth environments claimed by \citet{mar01b} (mostly on the basis
of the relatively small dimensions of the AXPs/SGRs supernova
remnants), has been   criticized and disproved \citep{dun02}.
Furthermore, the early suggestions for association with supernova
remnants for most AXPs/SGRs, on which this model is based, are no
more considered significant (also implying that no large
velocities are required, see sect. \ref{snr}).

Models involving fossil disks are often criticized based on the
fact that the putative disks should be visible in the optical and
NIR. The expected optical/IR flux   depends, among other things,
on the size and orientation of the disk, as well as on the
prescriptions assumed for the reprocessing  of the X-ray radiation
at longer wavelengths \citep{per00b,per00c,hul00,ert06b}. This
explains why different conclusions were drawn from such studies,
as is well exemplified by the case of \uu . This AXP is unique in
showing optical pulsations \citep{dhi05}, and has been detected
over a large wavelength range, from the B band to the mid-infrared
at 8 $\mu$m \citep{wan06}. When its optical counterpart was
identified, \citet{hul00} concluded that it was too faint to be
compatible with a disk, unless a particularly small disk size was
invoked. The subsequent discovery of optical pulsations at the
neutron star spin period \citep{ker02}, has been interpreted as
supporting the magnetar model, on the basis that the optical
(4,000-10,000 \AA) pulsed fraction (27\%) larger than the X-rays
one (\ltsima10\%) is difficult to explain in terms of
reprocessing. However, this argument assumes that the X-ray pulse
profile that we observe is the same of the radiation that
intercepts the disk, which might not be true due to orientation
and beaming effects. Recent observations of \uu\ with the Spitzer
Space Telescope revealed a mid-IR counterpart  at 4.5 and 8 $\mu$m
\citep{wan06}, interpreted as evidence for a cool (T$\sim$1000 K)
dust disk, truncated at an inner radius of $\sim3 \rsun$, and
non-accreting (i.e. a ''passive'' disk, heated by the magnetar
X-ray emission from the neutron star). On the other hand,
\citet{ert07} showed that both the mid and near IR fluxes and the
unpulsed optical emission are also consistent with an accretion
disk whose inner boundary is close to the corotation radius.

A more severe criticism to accretion-based models with residual
disks is that they cannot easily account for the bursts and
flares. Hence some additional mechanism has to be added in order
to explain these phenomena.

Finally, another possibility is that the magnetar field
responsible for the bursting activity is not dipolar, but it is
only present in higher order multipoles dominating near the
neutron star surface. In this ``hybrid'' scenario
\citep{ecs03,ert03}, the torque and accretion properties would be
determined by the interaction between the disk and a dipolar
component field, similar in strength to that of normal pulsars.

\subsection{Other models}
\label{exo}

Other models, not involving neutron stars,  have been put forward
in alternative to the magnetar and accretion models discussed
above. They are based on the possible existence of quark stars as
the most stable configuration for dense compact stars
\citep{xu07,hor07}.

Solid quark stars could emit bursts and giant flares powered by
gravitational energy released in star-quakes \citep{xu06}. Stars
made of strange quarks  in the ``color-flavor locked'' phase are
instead considered by \citet{ouy04}. The superconductive
properties of matter it this state determine how the surface
magnetic field adjusts itself to the internal field, which is
confined to the vortices. During this field alignment phase, the
star should be observable as a SGRs/AXPs.

P-stars made of up and down quarks in $\beta$-equilibrium with
electrons in a chromo-magnetic condensate have been suggested by
\citet{cea06}. This model still involves super strong dipolar
fields, but in P-stars rather than in neutron stars.

\section{RELATED OBJECTS}
\label{connections}

\subsection{Supernova remnants}
\label{snr}

Three of the nine confirmed AXPs, plus the candidate \axj , are
located at (or very close to) the geometrical center of shell like
supernova remnants (Table \ref{tab-list}).  The associations are
generally considered robust, due to the small chance probabilities
of these spatial coincidences.  Besides providing a way to obtain
the AXPs distances, this indicates that AXPs are young objects
(\ltsima 10$^{4}$ years) and do not have large transverse
velocities. The failure to detect SNR shells around the other
AXPs, despite targeted radio searches, is not in contradiction
with a small age for these objects. In fact, as also shown in the
case of several radio pulsars with small characteristic ages, the
remnants are not always visible, most likely owing to the
different conditions of the interstellar medium in the
surroundings of the supernova explosion.

\citet{gae01} critically examined the proposed SNR associations
for the four SGRs, which, if real, would imply large proper
motions for these neutron stars. They concluded that   only \lmc\
might be associated with a SNR, but the probability of a chance
coincidence for this SGRs lying close to the edge of N49, is not
as small as for the AXPs mentioned above.

\subsection{Massive star clusters}
\label{ass}

The relatively young ages of magnetars is also supported by a few
possible associations with clusters of massive stars. The
transient AXP \cxo\ was discovered during Chandra observations of
the open cluster Westerlund 1 \citep{mun06}. Clusters of massive
stars were also found close to the positions of \zerozero\
\citep{vrb00}, \zerosei\ \citep{eik01,fig05}, and \lmc\
\citep{klo04} during deep observations aimed at finding their
optical/IR counterparts. Although the chance probabilities of such
coincidences are difficult to estimate \textit{a posteriori}, it
is plausible that at least some of these objects were born in the
explosions of massive stars belonging to the clusters. The
projected separations between the magnetars and the cluster
centers are of $\sim$0.5--2 pc (except for \lmc , for which
d$\sim$30 pc). Considering the uncertainty in the ages, the
implied transverse velocities are consistent with  those of radio
pulsars.

The possible association of magnetars with star clusters is of
interest since it allows to set lower limits on the masses of
their progenitors, that must have evolved faster than the
currently observable cluster members. The young estimated ages of
Westerlund 1 (4$\pm$1 Myrs) and of the cluster close to \zerosei\
($<$4.5 Myrs) imply progenitors more massive than 40 $\msun$ and
50 $\msun$, respectively \citep{mun06,fig05}, while in the case of
the putative cluster of \zerozero\ the larger age ($<$10 Myrs)
gives a lower limit of only 20 $\msun$.

\subsection{Other classes of Isolated Neutron Stars}
\label{ins}

Observations in the X--ray, $\gamma$-ray and optical/IR bands have
significantly changed the old paradigm of isolated neutron stars
based mainly on the observations of the large population of radio
pulsars. Different new manifestations of isolated neutron stars,
besides AXPs and SGRs, have been recognized. Their existence might
simply reflect a larger variety in the birth properties of neutron
stars than previously thought, but it is also possible that some
of these classes of neutron stars are linked by evolutionary
paths.

The X-ray Dim Isolated Neutron Stars\footnote{This name is not
particulary appropriate anymore, considering that many dimmer
neutron stars have been revealed after the discovery of this class
of sources with the ROSAT satellite in the '90s. Only seven XDINS
are known, hence the nickname of ``Magnificent Seven'' often used
for these neutron stars.} (XDINS) are nearby ($\sim$100 pc) X--ray
pulsars characterized by very soft thermal spectra with blackbody
temperatures in the range 40--110 eV, X--ray luminosity of
10$^{30}$--10$^{32}$ erg s$^{-1}$, faint optical counterparts
(V$>$25), and absence of radio emission (see \citep{hab07} for a
recent review). Thanks to the complete absence of non-thermal
emission and, in a few cases, the measurement of parallactic
distances, they are often considered ideal targets to infer the
neutron star size and atmospheric composition through detailed
modelling of their purely thermal emission. A possible relation
with the magnetars is suggested by the fact that all the XDINS
have spin periods in the 3--12 s range, and the period derivatives
measured for two of them are of the order of 10$^{-13}$ s
s$^{-1}$. These P and $\pdot$ values give characteristic ages of
$\sim$1--2 Myrs and magnetic fields of a few 10$^{13}$ G (assuming
dipole radiation braking). Magnetic fields in the
$\sim10^{13}-10^{14}$ G range are also inferred by the broad
absorption lines observed in the X--ray spectra of most XDINS,
independently from their  interpretation either as proton
cyclotron features or atomic transition lines.

The seven objects observed within a distance of a few hundreds
parsecs imply that the space density of XDINS is much higher than
that of the active magnetars.  XDINS could thus  be the descendant
of magnetars. Note that more distant XDINS cannot be observed
because their very soft X-ray emission is severely absorbed in the
interstellar medium.

Periods similar to those of the magnetars are also seen in the
Rotating Radio Transients (RRATs)  recently discovered in the
Parkes Multibeam Survey \citep{mcl06}. These neutron stars emit
short (2-30 ms) pulses of radio emission at intervals of minutes
to hours. Their rotation periods, ranging from  0.4 to 7 s, could
be inferred from the greatest common divisors of the time
intervals between bursts. RRATs might represent a galactic
population as large as that of active radio pulsars, that remained
undiscovered for a long time due to lack of radio searches
adequate to detect them. The pulsed X--rays detected from one of
these objects have a thermal spectrum (blackbody temperature
$\sim$0.14 keV) and are consistent with cooling emission
\citep{mcl07}. Period derivatives have been determined to date for
three RRATs. Only one of these objects has a rather high inferred
field B=5$\times10^{13}$ G, while the other two have B = 3 and 6
$\times10^{12}$ G, similar to normal radio pulsars. Thus their
relation, if any, with the magnetars is unclear.

The Compact Central Objects (CCOs) form a heterogeneous group of
X--ray sources unified by their location at the center of
supernova remnants and by the lack of radio detections
\citep{pav04,del08}. These properties are shared with some of the
AXPs, indicating a possible connection between magnetars and CCOs.
The presence of supernova remnants implies that these are very
young objects, maybe in an evolutionary stage preceding the
AXP/SGR phase. However, the two CCOs for which pulsations have
been determined do not support such a relation and rather indicate
that these neutron stars are born with initial parameters opposite
to those of magnetars. They have short spin periods  (0.424 s and
0.105 s) and undetectable spin-down  ($\pdot$ \ltsima 2.5
10$^{-16}$ s s$^{-1}$), yielding estimated magnetic fields smaller
than a few 10$^{11}$ G \citep{got07,hal07}. The resulting
characteristic ages exceed by orders of magnitude their true ages,
as inferred from the associated SNRs, implying that their initial
rotational periods were not too different from the current values.
The low magnetic field and long initial spin periods of these
objects might be causally related.

Similar P and $\pdot$ have not been found in all the other CCOs,
despite intensive searches, and it cannot be excluded that some of
them be magnetars. Suggestions in this sense have been done, e.g.,
for the CCOs in RCW 103 (discussed in the next section) and in Cas
A. Infrared features with apparently superluminal motion were
observed outside the shell of the Cas A supernova remnant and
interpreted as light echoes of a recent outburst from the CCO
\citep{kra05}. The geometry of two of such light echoes, at
opposite sides of the remnant, is consistent with the emission
from the CCO of a short pulse of radiation, beamed nearly
perpendicular to the line of sight. This energetic event,  that
should have occurred between 1950 and 1955, would have been
similar to an SGR giant flare, implying that the Cas A CCO is a
magnetar. However, a recent analysis of more IR data failed to
confirm this scenario and indicates that all the light echoes
surrounding the remnant can be traced back to the date of the Cas
A supernova explosion \citep{kim08}.

The most sensitive radio surveys carried out in the last decade
have extended the range of observed  magnetic fields\footnote{as
inferred from the timing parameters with the usual dipole
assumption.} in rotation powered pulsars, leading to the discovery
of a few objects with fields approaching those of magnetars.
However, no signs of magnetar-like activity, such as enhanced
X-ray emission or bursts,  were seen in the rotation-powered radio
pulsars with the highest inferred magnetic fields (several
10$^{13}$ G) \citep{cam00,mcl04}. For example, PSR J1814--1744,
despite having P and $\pdot$ values very similar to those of the
AXP \ee\ has a 2-10 keV  luminosity smaller than 2 10$^{33}$ erg
s$^{-1}$ \citep{piv00}. These findings seemed to indicate that the
dipole magnetic field intensity was not by itself the only element
responsible for differentiating magnetars from ordinary radio
pulsars. Very recently, short bursts have been discovered from the
young pulsar  at the center of the Kes 75 supernova remnant
\citep{gav08b}. This object, PSR J1846--0258 (P=0.326 s), is the
pulsar with the smallest known characteristic age (884 yrs) and
has a high field of 5 10$^{13}$ G. Its lack of radio emission was
generally ascribed to beaming, but the discovery  of magnetar-like
activity now leads to consider also the possibility that this
pulsar be truly radio silent. The bursts observed in  PSR
J1846--0258 are very similar to those seen in AXPs, and are
accompanied by an enhancement of the persistent X--ray emission, a
spectral softening and an increased timing noise \citep{gav08b}.
The important discovery that apparently normal rotation-powered
pulsars can exhibit the same kind of magnetically driven activity
seen in AXPs and SGRs points to a more strict connection between
radio pulsars and magnetars than previously thought.

\subsection{The CCO in RCW 103: a braked down magnetar ?}
\label{rcw}

\begin{figure*}
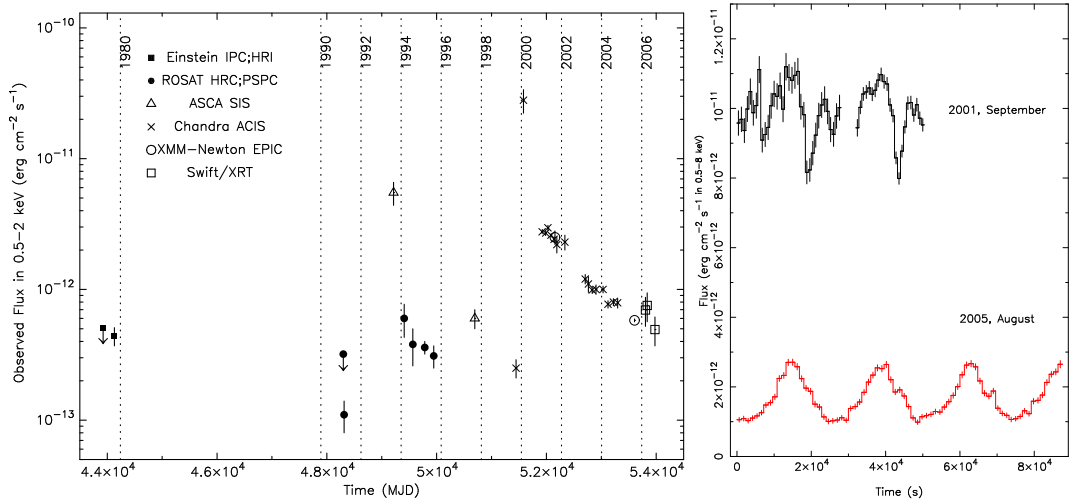

\hbox{
 \psfig{figure=rcw103_lc_del06.ps,angle=-90,width=9cm}
 \psfig{figure=rcw103_pp_del06.ps,angle=0,width=5cm}
} \caption{Long term X--ray light curve (left panel) and pulse
profiles (right panel) of the CCO in RCW 103 (from \citet{del06}).
Note the different pulse profiles corresponding to the two
different source intensity states.  }
 \label{fig-rcw103}
\end{figure*}

The X--ray source 1E 161348--5055 in the supernova remnant RCW 103
has unique variability properties that clearly distinguish it from
the other CCOs \citep{del06}. It showed secular luminosity
variations in the range 10$^{33}$--10$^{35}$ erg s$^{-1}$ and its
flux is strongly modulated with a period of 6.7 hours
(Fig.~\ref{fig-rcw103}). No faster periods have been detected. The
X--ray pulsed fraction larger than 40\%, the light curve
variability, and the optical/NIR limits, ruling out companion
stars of spectral type earlier than M5, exclude the interpretation
of the 6.7 hr modulation as the orbital period of a normal low
mass X-ray binary (a possibility that is also ruled out by the
young age, $\sim$2 kyrs, of RCW 103). It seems thus more likely
that the periodicity is due to the slow rotation of an isolated
neutron star and that the observed X--ray emission is magnetically
powered. In this scenario one is faced with the problem of slowing
down the magnetar to such a long rotation period within the short
lifetime of only a few thousand years implied by the age of the
RCW 103 supernova remnant. A viable possibility \citep{del06} is
that the braking was provided by the propeller effect due to the
presence of a fossil disk formed from the supernova material
fall-back. This evolutive path requires a neutron star initial
period longer than $\sim$300 ms in order to avoid the disk
disruption by the relativistic outflow of the newly born active
radio pulsar. If this interpretation is correct, it would support
other recent evidence that high magnetic fields might also be
present in NS born with long spin periods \citep{fer06}, contrary
to the standard magnetar formation scenario  discussed in section
~\ref{formation}.

Alternatively, the RCW 103 CCO could be a binary formed by a very
low mass star and a magnetar with a spin (quasi-)synchronous with
the orbital period \citep{piz08}. In this model the torque needed
to slow down the neutron star can be provided by magnetic and/or
material interactions, similar to the case of white dwarfs in
intermediate polars.

\subsection{Gamma-ray bursts}
\label{grb}

The initial short spikes characterizing SGR giant flares have
spectral and duration properties similar to those of the
short-hard class of gamma-ray bursts. It was therefore suggested
that short gamma-ray bursts could be giant flares from SGRs in
distant galaxies, for which only the bright initial peaks can be
seen while the following pulsating tails remain below the
detection sensitivity. Although this idea dates back to the time
of the SGRs discovery (e.g. \citep{maz82}), it received renewed
attention after the 2004 giant flare from \zerosei , due to the
high peak luminosity of this event (a factor 100 larger than that
of the two previously observed giant flares, if \zerosei\ is
indeed at 15 kpc, see Table \ref{tab-gf}).

Short bursts ($<$2 s) made up about one quarter of the bursts
detected by BATSE, with an all-sky detection rate of $\sim$170
yr$^{-1}$, while in other satellites, more sensitive in a lower
energy range, they constitute a smaller fraction of the total GRB
sample. A giant flare like that of \zerosei\ would have been
visible by BATSE within a distance D$_{max}$ =
($\frac{F_{1806}}{F_{thr}})^{1/2}$ D$_{1806}$, where F$_{1806}$ is
the flare peak flux (or fluence) and F$_{thr}$ the assumed trigger
threshold for BATSE. For D$_{1806}$=15 kpc, and considering the
uncertainties in the above values,  one obtains D$_{max}\sim$30-50
Mpc.  The expected rate of detectable SGR flares within this
volume depends on further quantities not known very precisely: the
frequency of giant flares in our Galaxy, estimated from only three
events, and the star formation rate in the local universe compared
to that in the Galaxy. In fact it is reasonable to assume that
SGRs, being young objects associated with massive stars, have an
abundance proportional to the star formation rate. Owing to all
these factors, different estimates, ranging from $\sim$40-50\%
\citep{hur05,pal05}, up to 100\% \citep{nak06}, were obtained for
the fraction of short bursts in the BATSE sample that could be due
to extragalactic SGRs.

However, these optimistic estimates are contradicted by several
observations and analysis. No excess of short BATSE bursts is
found in the direction of the Virgo cluster (d$\sim$17 Mpc), nor
short bursts were consistent with the direction of the closest
galaxies with a high star formation rate \citep{pal05,pop06}.
Searches in the error regions of a few well localized short bursts
failed to detect nearby galaxies \citep{nak06}. A spectral
analysis of a sample of short BATSE bursts showed that only a
small fraction are spectrally consistent with a SGR flare
\citep{laz05}. Finally, the Swift/BAT instrument, being able to
detect a \zerosei -like flare up to $\sim$70 Mpc \citep{hur05},
should have observed a larger number of short bursts. All these
findings suggest that some of the above assumptions are not valid.
One possibility is that the distance of \zerosei\  be lower than
15 kpc \citep{bib08}. More likely, the assumed galactic rate of
one giant flare every $\sim$30 years per source does not apply to
the most energetic flares. Obviously these conclusions do not
exclude the possibility that some of the short GRBs be due to
SGRs, and in fact a few candidates have been reported in recent
years (see Table~\ref{tab-short}).


\begin{table}[t]
\caption{Candidate extragalactic SGRs} \centering
\label{tab-short}       
\begin{tabular}{lcccccc}
\hline\noalign{\smallskip}

GRB &  Galaxy  & distance &  Duration &  Energy     & Comments  & Ref.  \\
     &         &  (Mpc)   & (s)      &  (ergs)      &   &  \\

\tableheadseprule\noalign{\smallskip}
 970110 & NGC 6946 ?&  5.9   & 0.4  &  2.7 10$^{44}$  &  possible P=13.8 s in tail  & \cite{cri06}   \\
 000420B&  M74      &  10.4  & 0.3  &  3 10$^{46}$    &   & \cite{ofe07}  \\
 051103 &  M81      &  3.6   & 0.2  &  7 10$^{46}$    &  &  \cite{ofe06,fre07b}   \\
 070201 &  M31      &  0.78  & 0.15 &  1.5 10$^{45}$  &   & \cite{ofe08,maz08} \\

\noalign{\smallskip}\hline
\end{tabular}
\end{table}

Gamma--ray bursts might also be associated to the formation of
magnetars. Rapidly rotating, ultra-magnetized proto-neutron stars
can provide the central engine required to sustain for a
sufficiently long time the observed emission \citep{buc08}.
\citet{met08} propose that the short GRBs with extended soft
emission, like GRB 060614 \citep{geh06} are produced by
proto-magnetars formed in accretion induced collapse of white
dwarfs or in the merging of white dwarf binaries. The extended
emission lasting 10-100 s observed after these short bursts would
result from a relativistic wind powered by the proto-magnetar
rotational energy.

\section{CONCLUSIONS AND FUTURE PROSPECTS}
\label{future}

We can expect that, as usual for astronomical objects with extreme
properties, the  interest in AXPs and SGRs will  not decrease in
the coming years. In the immediate future, the prospects for large
advances in the classical X-ray band are somehow limited, owing to
the paucity of new missions significantly improving the
capabilities of the currently available big observatories like
XMM-Newton and Chandra. RXTE will soon stop operations, after a
long and very successful series of observations that will be
difficult to equal for what concerns all the timing aspects.
XMM-Newton and Chandra have already provided a good harvest of
data on most magnetars, but it is important to continue these
observations especially in view of the variability phenomena
discussed above. These data will remain for many year the basic
reference for all the spectral models now being developed trying
to include a realistic treatment of the physical processes and
conditions in magnetars. Of course, new transients, as well as
outbursts/flares from the known magnetars, hold the greatest
potential for interesting discoveries.

The situation is more promising for what concerns the hard X--ray
band, where the presence of significant emission from most
magnetars has been established, but detailed studies are hampered
by the relatively poor sensitivity of the current instruments. A
few satellite missions  (NuStar, Simbol-X, Next) are now being
developed and expected to be operational after 2012. They will
provide a significant step forward in sensitivity thanks to the
introduction of X--ray focussing at hard X-ray energies.

On a more immediate time frame, the gamma-ray band above 100 MeV
can be explored with AGILE and GLAST. In regions close to the
magnetars polar caps, the strong magnetic field is expected to
quench gamma-ray emission due to pair production. However, the
magnetic field is much lower in the regions considered for the
gamma-ray production in outer gap models for radio pulsars. The
application of such models to magnetars leads to predicted
gamma-ray flux above the expected GLAST sensitivity \citep{che01}.
Positive detections and phase resolved spectral studies in the MeV
range would provide important comparison with rotation powered
gamma-ray pulsars allowing to test the models over a wider range
of the relevant parameters.

Relativistic baryons accelerated in giant flares make the  SGR
potential sources of neutrinos \citep{iok05} and high-energy
cosmic rays \citep{asa06}. The AMANDA-II neutrino detector gave
only an upper limit for the \zerosei\ December 2004 giant flare
\citep{ach06},  and also a search for ultra high energy cosmic
rays associated to this event gave a negative result
\citep{anc07}. However, future experiments might well confirm
these prediction if other suitable giant flares are observed.
Furthermore,  ultra high-energy cosmic rays could be produced in
the relativistic winds of rapidly spinning  magnetars immediately
after their birth \citep{aro03}.

The seismic vibrations which are thought to be at the origin of
the QPOs seen in giant flares (Sect.~\ref{qpo}) also  produce
gravitational waves.  A search in the LIGO data at the frequencies
seen in the \zerosei\ giant flare provided significant upper
limits \citep{abb07}. The emission of gravitational waves is also
expected during the formation of magnetars. In fact, if events as
powerful as the December 2004 giant flare are not unique in a
magnetar lifetime, energetic arguments\footnote{Assuming for
\zerosei\ a distance of 15 kpc.} require that the internal
magnetic field of a newly born magnetar be larger than 10$^{16}$
G. Such a high field can induce a substantial deformation in the
neutron star, which can give rise to the emission of gravitational
waves if the rotation and symmetry axis are not aligned
\citep{ste05}. Thus, there is the exciting possibility that
magnetars might be among the first detected sources of
gravitational waves.

\begin{landscape}
\begin{table}[t]
\caption{Coordinates (J2000) of AXPs and SGRs}
 \centering
\label{tab-positions}

\begin{tabular}{lcccc}
\hline\noalign{\smallskip}
Name &  X-ray position$^{(a)}$  & Uncertainty$^{(c)}$  & Counterparts &    \\[3pt]
     &                  & (arcsec)  &              &  \\[3pt]

\tableheadseprule\noalign{\smallskip}

\smcl        & 01 00 43.03    & 0.5  (1$\sigma$?)        &   & IR candidate   \\
             & --72 11 33.6    &    \cite{mcg05}    &   &      \\
 \hline
\uul         & 01 46 22.44     & 0.5 (1$\sigma$?) & 01 46 22.41   & opt. counterpart \\
            &  +61 45 03.3   &   \cite{jue02} & +61 45 03.2  & \cite{hul00}     \\
\hline
\ool         & 10 50 07.14     &  0.6 (90\% c.l.)    & 10 50 07.13 &  opt. counterpart   \\
            & --59 53 21.4     & \cite{wan02}    &  --59 53 21.3   & \cite{wan02}       \\
\hline
\quil      &  15 50 54.11     & 0.8 (99\% c.l.)   &  15 50 54.11   & 0.1 radio     \\
         &   --54 18 23.8    & \cite{gel07}  &  --54 18 23.7   & \cite{cam07c}     \\
\hline
\cxol    &  16 47 10.2 &  0.3 (90\% c.l.)   &     &     \\
        &  --45 52 16.9 &   \cite{mun06}  &     &      \\
\hline
\rxsl     & 17 08 46.87   & 0.7 (90\% c.l.)   &    &  \\
        & --40 08 52.44   &     \cite{isr03} &   &         \\
\hline
\xtel     & 18 09 51.08   & 0.6 (90\% c.l.)   & 18 09 51.087  & radio   \\
       &  --19 43 51.7   & \cite{got04}       & --19 43 51.93  & \cite{cam07b}   \\
\hline
\kesl     & 18 41 19.343   &  0.3 (1$\sigma$)   &    &  \\
         & --04 56 11.16  &   \cite{wac04}  &     &       \\
\hline
\axjl     &  18 44 57  & 120$^{(b)}$ (90\% c.l.)    &   18 44 54.68  &  0.6 (90\%) possible Chandra ctpt    \\
        &  --03 00   & \cite{tor98}    &  --02 56 31.1   &  \cite{tam06}  \\
\hline
\eel      & 23 01 08.295   &  0.6 (99\% c.l.)     & 23 01 08.312     &  NIR ctpt   \\
         & +58 52 44.45  & \cite{hul01}           &  +58 52 44.53   &  \cite{hul01}      \\
\hline
 \hline
\lmc     &  05 26 00.89 & 0.6  (1$\sigma$) &   &    \\
       &  --66 04 36.3  & \cite{kul03}   &  &      \\
\hline
\sedici  & 16 35 51.844     & 0.2 (1$\sigma$)  &      & \\
        &  --47 35 23.31     &  \cite{wac04}          &      & \\
\hline
\zerosei & 18 08 39.32   &  0.3 ($1\sigma$)  & 18 08 39.337   & NIR ctpt  \\
         & --20 24 39.5   &  \cite{kap02b}  &  --20 24 39.85  & \cite{isr05}     \\
\hline
\zerozero&         &         &     19 07 14.33 & 0.15 radio      \\
         &         &         &     +9 19 20.1 & \cite{fra99}     \\
\hline \hline

\noalign{\smallskip}\hline
\end{tabular}

Notes:

(a) all the positions are from Chandra observations, except for
\axj\

(b) position obtained with ASCA

(c)  confidence levels of the error radii are given hers as
reported in the corresponding references. A question mark
indicates that the confidence level was not explicitly given.

\end{table}
\end{landscape}



\bibliographystyle{aa}
\bibliography{axpsgr}   

\end{document}